\documentclass[twocolumn,prl,superscriptaddress]{revtex4-1}
\usepackage{amsmath,amssymb,mathrsfs}
\usepackage{natbib}
\usepackage{subfigure}
\usepackage{tabularx}
\usepackage{epsfig}
\usepackage{longtable}
\usepackage{amsfonts}
\usepackage{rotating}
\usepackage{bbold}
\usepackage{hhline}
\usepackage{braket}
\usepackage{txfonts, comment}
\usepackage{multirow}
\usepackage{appendix}
\usepackage[unicode=true,bookmarks=true,bookmarksnumbered=false,bookmarksopen=false,breaklinks=false,pdfborder={0 0 1},backref=false,colorlinks=true]{hyperref}

\hypersetup{linkcolor=magenta,urlcolor=blue,citecolor=blue,pdfstartview={FitH},hyperfootnotes=false,unicode=true}

\def\be{\begin{equation}}
\def\ee{\end{equation}}
\def\bea{\begin{eqnarray}}
\def\eea{\end{eqnarray}}

\begin{document}

\title{Statistical Symmetry Breaking and Emergent Colored Noise in a Stochastic Scalar-Doublet Field Theory}

\author{Pei Wang}
\affiliation{Department of Physics, Zhejiang Normal University, Jinhua 321004, China}
\email{wangpei@zjnu.cn}

\begin{abstract}
We investigate a relativistic stochastic field theory in which a complex scalar doublet is coupled to a complex white-noise source. The action preserves Lorentz and $\mathrm{U(1)}\times\mathrm{SU(2)}$ symmetries at the statistical level, whereas the corresponding Euler-Lagrange equations exhibit symmetry breaking along individual stochastic realizations. Within a gauge-field-free sector introduced to obtain analytical solutions, we show that the scalar doublet undergoes a noise-driven random walk in field space, leading to a finite, time-dependent ensemble average of its magnitude. As an illustrative application, we further investigate the coupling of the stochastic scalar field to fermions through a Yukawa interaction. The scalar-field solution naturally separates into a tail component, which contributes as an effective mass-like term, and a light-cone component, which acts as a colored-noise source that induces a spatially correlated stochastic phase in the fermion wave function. The statistical properties and correlation length of this emergent colored noise are derived analytically within the adopted approximations. The present work provides an exploratory study of statistical symmetry breaking and emergent colored-noise dynamics in a relativistic stochastic scalar-doublet field theory.
\end{abstract}

%\date{\today}

\maketitle

\section{Introduction}
\label{sec:intro}

Stochastic quantum-state dynamics has been actively studied as a possible framework
for explaining wave-function collapse and the intrinsic randomness of measurement
outcomes in quantum foundations~\cite{GRW,Diosi89,CSL,CSL2,Penrose96,Pearle99,Bassi05,Adler07,Adler08,Bassi13,Vinante16,Vinante17,Bahrami18,Tilloy19,Pontin19,Vinante20,Zheng20,Komori20,Donadi21,Gasbarri21,Carlesso22}. In these theories, both the unitary evolution and
the collapse of the wave function are treated as objective physical processes. The
quantum state is assumed to follow a random trajectory in Hilbert space, governed
by a stochastic differential equation that generalizes the 
Schr\"odinger equation~\cite{GRW,Diosi89,CSL,CSL2,Penrose96,Pearle99,Bassi05,Adler07,Adler08,Bassi13}.
Since the fundamental laws of modern physics are formulated in the language of
quantum field theory (QFT), most notably the Standard Model, it is natural to ask
how stochastic quantum-state dynamics can be incorporated into a relativistic
field-theoretic framework. Such an incorporation leads to stochastic quantum field
theories (SQFTs), which describe random quantum-state evolution while respecting
Lorentz symmetry~\cite{Wang22}.

Directly constructing Lorentz-invariant stochastic differential equations,
however, is technically challenging~\cite{Pearle99,Bassi13,Myrvold17,Tumulka20,Jones20,Jones21}. 
To address this difficulty, we recently
developed an action-based approach to SQFT, in which random-valued scalar terms are
added to the conventional QFT action~\cite{Wang22,Wang24}. These terms are constructed from quantum
fields---scalar, vector, or spinor---that represent matter, together with
Lorentz-invariant noise fields representing external stochastic influences that
drive the random evolution. The noise field may be a white noise $dW(x)$ or a more
general colored noise constructed using 1+3-dimensional stochastic calculus, and
is required to remain statistically invariant under Lorentz transformations. An
important advantage of this approach is that it can be straightforwardly
integrated into existing particle-physics models. By introducing only a small
number of noise terms into the Standard Model action, one may hope to describe
quantum-state collapse while leaving the established properties of elementary
particles essentially unchanged in the weak-noise limit.

White noise has a vanishing correlation length and is thus inappropriate for 
direct coupling to density operators, which are widely believed to couple instead to 
colored noise with a finite correlation length~\cite{Bassi13}. To model fermionic
wave-function collapse, we previously proposed an action~\cite{Wang24} in which the fermion
density operator $\bar{\psi}\psi$ is coupled to a Lorentz-invariant colored noise
field $h(x)$. The field $h(x)$ is obtained by solving the d'Alembert equation with
$dW(x)$ as a source. We showed that the correlation of $h(x)$ decays exponentially
with spatial separation and derived an analytical expression for its correlation
length. Although this model may serve as a potential candidate for describing
wave-function collapse, a rigorous derivation of Born's rule remains an open
challenge.

On the other hand, the use of a relatively complicated colored noise makes the
previous model less compelling as a fundamental theory than approaches based
directly on the simpler white noise $dW(x)$. Since the colored noise $h(x)$
arises as a solution of the d'Alembert equation, it is natural to ask whether
such colored noise can emerge dynamically from a relativistic stochastic field
theory in which a scalar field is coupled directly to white noise. This
observation motivates us to investigate a stochastic field theory of a complex
scalar doublet interacting with a statistically invariant white-noise source.
The scalar doublet is chosen because it possesses the same
$\mathrm{U(1)}\times\mathrm{SU(2)}$ internal symmetry as the electroweak scalar
sector and admits Yukawa couplings to fermions, thereby providing a convenient
prototype for studying how colored noise may emerge and subsequently influence
fermion dynamics.

This paper is devoted to the study of this stochastic scalar-doublet model.
Our primary objective is to investigate the mathematical properties of the
resulting stochastic field equations, including statistical symmetry breaking,
the emergence of a nonzero ensemble average of the scalar-field magnitude, and
the appearance of colored noise through the solution of the field equation.
From this perspective, the present work may also be viewed as a mathematical 
investigation of Lorentz-invariant stochastic differential equations in 1+3 dimensions, 
illustrating how statistical symmetry breaking and colored stochastic structures can 
emerge naturally from their solutions.
The Yukawa coupling to fermions is then considered as an illustrative
application for exploring how the emergent colored noise influences fermionic
wave functions. Although the scalar doublet considered here possesses the same internal
$\mathrm{U(1)}\times\mathrm{SU(2)}$ symmetry as the electroweak scalar sector,
the present work should be regarded as an exploratory toy model rather than a
complete stochastic extension of the Standard Model. In
particular, we do not attempt to reproduce the full phenomenology of the Higgs
sector or the coupled dynamics of the electroweak gauge fields.

The stochastic scalar-doublet model studied here is also of potential interest
from a cosmological perspective. Scalar fields with the same internal symmetry
structure as the electroweak Higgs sector have been widely investigated in
cosmology, including studies of Higgs vacuum metastability, Higgs-driven
inflation, and spacetime-dependent scalar expectation values induced by
gravitational effects~\cite{JHEP08_2012,PRD82_065008,Ref2103,Ref2307,Ref2411}.
The stochastic mechanism considered in the present work differs from these
approaches by introducing statistical fluctuations through coupling to white
noise rather than through a classical scalar potential or gravitational
effects. Whether such stochastic dynamics has observable cosmological
consequences remains an interesting question for future investigation.

The remainder of this paper is organized as follows.
In Sec.~\ref{sec:model}, we introduce the stochastic scalar-doublet model,
explain the physical meaning of its constituent terms, and clarify the
statistical symmetries it respects.
In Sec.~\ref{sec:EL}, we derive the Euler-Lagrange equation governing the
real-time dynamics of the scalar doublet and construct its explicit solution.
Section~\ref{sec:Htt} is devoted to the analysis of this solution. We show that
the noise-driven dynamics induces a random walk of the scalar doublet in the
complex field space, leading to statistical symmetry breaking and a nonzero
ensemble average of the field magnitude.
In Sec.~\ref{sec:fdH}, we investigate an illustrative application in which the
scalar doublet is coupled to fermions through a Yukawa interaction. We show
that the resulting solution gives rise to two distinct effects: an effective
mass-like contribution and an emergent colored noise acting on fermionic
degrees of freedom, whose statistical properties are analyzed in detail.
In Sec.~\ref{sec:para}, we provide rough estimates of the free parameters of
the model and verify the self-consistency of the approximations employed
throughout the paper.
Finally, Sec.~\ref{sec:conclusion} summarizes our main results and discusses
their limitations together with several directions for future work.

\section{Action and symmetry}
\label{sec:model}

We consider a complex scalar doublet $H(x) = \bigl( H_A(x),\, H_B(x) \bigr)^T$, 
where each component can be decomposed as
$H_{\alpha} = H_{\alpha R} + i H_{\alpha I}$ with $\alpha = A, B$.
Here, $H_{\alpha R}$ and $H_{\alpha I}$ denote the real and imaginary parts,
respectively. Throughout this work, $H(x)$ is studied as a
stochastic scalar-doublet field. It carries the same internal
$\mathrm{U(1)}\times\mathrm{SU(2)}$ quantum numbers as the Standard
Model Higgs doublet, and the model is therefore motivated by the Higgs
sector. However, the analysis presented here is intended as a
field-theoretic toy model rather than as a complete description of the
electroweak Higgs sector. 

The scalar doublet is coupled to a complex doublet white-noise field
\begin{equation}
\begin{split}
d\Omega(x)
= \begin{pmatrix}
d\Omega_A(x) \\
d\Omega_B(x)
\end{pmatrix},
\end{split}
\end{equation}
where $d\Omega_A(x)$ and $d\Omega_B(x)$ represent two independent complex white-noise
fields. Each component can be written as
\begin{equation}
d\Omega_{\alpha}(x)
= dW_{\alpha R}(x) + i\, dW_{\alpha I}(x),
\end{equation}
with $\alpha = A, B$. Here, $dW_{AR}$, $dW_{AI}$, $dW_{BR}$, and $dW_{BI}$ 
are four independent real white-noise fields.

The real white-noise field $dW(x)$ and its properties were introduced previously~\cite{Wang22, Wang24}.
It is defined by partitioning spacetime into infinitesimal cells of volume $d^4 x$,
and assigning to each cell a random number drawn from a Gaussian distribution with
zero mean and variance $d^4 x$. The fundamental properties of $dW(x)$ are
$\bigl(dW(x)\bigr)^2 = d^4 x$, while $dW(x)dW(x')$ for $x \neq x'$ can be neglected,
as dictated by stochastic calculus~\cite{Mikosch98app,Wang22,Wang24}. For the complex noise introduced here, each
spacetime point $x$ is associated with four independent identically distributed
(i.i.d.) real random variables,
$dW_{AR}(x)$, $dW_{AI}(x)$, $dW_{BR}(x)$, and $dW_{BI}(x)$. Noise variables at
different spacetime points are independent, forming a collection of independent
Gaussian random numbers.

A natural scalar coupling between $H(x)$ and $d\Omega(x)$ is given by either
$H^\dag d\Omega$ or $d\Omega^\dag H$. To ensure that the action is real, we choose
the symmetric combination
\begin{equation}
\label{eq:action:co}
\begin{split}
H^\dag d\Omega + d\Omega^\dag H
= 2 \Big(
& H_{AR}\, dW_{AR} + H_{AI}\, dW_{AI} \\
& + H_{BR}\, dW_{BR} + H_{BI}\, dW_{BI}
\Big).
\end{split}
\end{equation}
In Ref.~[\onlinecite{Wang22}], we studied the coupling of a single real scalar field to white noise.
Equation~\eqref{eq:action:co} represents a straightforward generalization to the
case of multiple field components.

In flat Minkowski spacetime, the coupling in
Eq.~\eqref{eq:action:co} is statistically invariant under spacetime translations
and Lorentz transformations. However, in curved spacetime or in generic coordinate
systems, this expression is not a scalar but a scalar density. To construct a
genuine scalar quantity, an additional factor of $\bigl(-g\bigr)^{1/4}$ must be
included, where $g = \det(g_{\mu\nu})$ is the determinant of the metric tensor. This
requirement originates from the fact that the variance of $dW(x)$ is $d^4 x$, which
is not invariant under general coordinate transformations, whereas the spacetime
volume element $\sqrt{-g}\, d^4 x$ is invariant.
Consequently, only the combination $\bigl(-g\bigr)^{1/4} dW(x)$, which
has variance $\sqrt{-g}\, d^4 x$, possesses the appropriate transformation properties.

We emphasize that our analysis is carried out in curved spacetime, in particular
in an expanding spacetime with a finite initial time (the Big Bang). This choice
is not only motivated by cosmological realism, but also by mathematical
consistency: the theory is free of divergences and thus well defined only in such
spacetimes. By contrast, in flat Minkowski spacetime the model exhibits an
infrared divergence, as will be demonstrated below.

We construct a random-valued scalar action of the form
\begin{equation}
\label{eq:action:ido}
\begin{split}
& \int \bigl(-g(x)\bigr)^{1/4}
\left( H(x)^\dag d\Omega(x) + d\Omega(x)^\dag H(x) \right) \\ = & \, 
2 \int \bigl(-g(x)\bigr)^{1/4}  \Big(
 H_{AR}(x) \, dW_{AR}(x)  + H_{AI}(x) \, dW_{AI} (x) \\ & 
+ H_{BR}(x) \, dW_{BR}(x) + H_{BI}(x) \, dW_{BI}(x)\Big).
\end{split}
\end{equation}
For completeness, we briefly summarize the definition of the stochastic
integral employed throughout this work. For a
function $f(x)$, the stochastic integral $\int dW(x) \, f(x)$ is defined
as follows. Given a partition of spacetime into four-dimensional cells
of volume $\Delta^4 x_i$, independent Gaussian random variables
$\Delta W(x_i)$ are assigned to each cell, with vanishing mean and
variance $\Delta^4 x_i$. The stochastic integral is then defined as the
continuum limit $\int dW(x)\, f(x)
\equiv \lim_{\max(\Delta^4 x_i)\to 0}
\sum_i f(x_i)\,\Delta W(x_i)$. In our previous
works~\cite{Wang22,Wang24}, we proved that this limit is well defined,
independent of the particular choice of spacetime partition, and yields
a unique stochastic integral. Since the focus of the present work is on
the physical implications of the stochastic scalar-doublet model, we refer
interested readers to these references for the detailed mathematical
construction and proofs.

Equation~\eqref{eq:action:ido} defines a random-valued scalar in the sense that it
is statistically invariant under general coordinate transformations. More
explicitly, under a transformation $x \to x'$, the scalar-doublet field transforms as
$H'(x') = H(x)$, while the statistical properties of
$\bigl(-g(x)\bigr)^{1/4} d\Omega(x)$ remain unchanged. As a result, the quantity
in Eq.~\eqref{eq:action:ido}, evaluated in different coordinate systems, has
exactly the same probability distribution.

We stress that, in stochastic quantum field theory, the action is itself a
random-valued quantity. Consequently, the deterministic symmetry principle of
ordinary quantum field theory---requiring the action to be strictly invariant
under transformations---must be replaced by a statistical symmetry principle,
which only requires invariance of the probability distribution of the action.

The complete action for the scalar-doublet field in a general curved spacetime is then
given by
\begin{equation}
\label{eq:action}
\begin{split}
S = {} &
- \int d^4 x \sqrt{-g}\, g^{\mu\nu} D_\mu H^\dag D_\nu H
- \eta^2 \int d^4 x \sqrt{-g}\, H^\dag H
\\ &
+ \gamma \int \bigl(-g\bigr)^{1/4}
\left( H^\dag d\Omega + d\Omega^\dag H \right),
\end{split}
\end{equation}
where $D_\mu = \partial_\mu- \frac{i g}{2} W_\mu^a \sigma^a - \frac{i g'}{2} B_\mu$
is the gauge-covariant derivative, with $W_\mu^a$ and $B_\mu$ denoting the gauge
boson fields, $\sigma^a$ the Pauli matrices, and $g$ and $g'$ the corresponding
gauge coupling constants. The model contains two free parameters: $\gamma$,
which controls the coupling strength between the scalar-doublet field and the white noise,
and $\eta$, which parametrizes a mass term. Throughout this paper, we
adopt the metric signature $(-,+,+,+)$ and set $\hbar = c = 1$.

As discussed above, the first two terms in the action~\eqref{eq:action} are
strictly invariant under arbitrary coordinate transformations, while the third,
random-valued term is invariant in the statistical sense. The full action
therefore respects statistical Lorentz symmetry and statistical spacetime
translation invariance.

Let us clarify the mass dimensions of the relevant quantities. We adopt the
convention that the scalar-doublet field has mass dimension one, or equivalently dimension
$\mathrm{L}^{-1}$. The noise field $d\Omega(x)$ must then have dimension
$\mathrm{L}^2$, since its variance is proportional to $d^4 x$, which has
dimension $\mathrm{L}^4$. Consequently, the parameters $\eta$ and $\gamma$ must
carry the same dimension, namely $\mathrm{L}^{-1}$, corresponding to a mass
dimension of one.

The stochastic scalar-doublet model considered here is motivated by the
electroweak Higgs sector but is not intended as a complete model of Higgs
physics. The scalar doublet carries the same
$\mathrm{U(1)}\times\mathrm{SU(2)}$ quantum numbers as the Standard
Model Higgs field and is coupled to fermions through a Yukawa interaction,
allowing us to investigate how stochastic scalar-field dynamics may
influence fermionic degrees of freedom. However, the conventional Higgs
potential,
$\mu^2 H^\dag H-\lambda(H^\dag H)^2$,
is replaced by a simplified mass term,
$-\eta^2H^\dag H$,
together with a stochastic coupling to white noise.
The purpose of this construction is therefore not to reproduce the full
phenomenology of the Standard Model Higgs sector. Rather, it serves as a
field-theoretic toy model for studying the consequences of stochastic
dynamics in a complex scalar doublet. In particular, we investigate
whether noise-driven evolution can generate a nonzero ensemble average of
the scalar-field magnitude, $\langle|H|\rangle$, and how the resulting
scalar-field configuration induces an effective mass-like contribution and
an emergent colored stochastic noise through the Yukawa interaction. The
extent to which these ideas can be incorporated into a realistic
electroweak theory capable of reproducing the observed Higgs boson
properties remains an open question and is left for future work.

In addition to statistical Lorentz and spacetime translation symmetries, the
action~\eqref{eq:action} also exhibits an important
$\mathrm{U}(1) \times \mathrm{SU}(2)$ symmetry. A general
$\mathrm{U}(1) \times \mathrm{SU}(2)$ transformation can be written as
\begin{equation}
\mathcal{U}(x)
= \exp\!\left[
\frac{i}{2} \theta^a(x) \sigma^a
+ \frac{i}{2} \theta^0(x)
\right],
\end{equation}
where $\theta^a(x)$ ($a=1,2,3$) and $\theta^0(x)$ are arbitrary real functions of
spacetime. Under such a transformation, the scalar-doublet field transforms as
$H(x) \to H'(x) = \mathcal{U}(x) H(x)$, and the gauge-covariant derivative transforms
accordingly as
$D_\mu H(x) \to D'_\mu H'(x) = \mathcal{U}(x) D_\mu H(x)$. The first two
deterministic terms in the action~\eqref{eq:action} are therefore strictly
invariant under $\mathcal{U}(x)$.

The complex doublet noise field $d\Omega(x)$, on the other hand, represents an
external stochastic potential acting on matter fields. It is defined solely by
the spacetime partition, in particular by the cell volume $d^4 x$, and must
therefore be independent of the choice of $\mathcal{U}(x)$. To analyze the
symmetry properties of the coupling $H^\dag d\Omega$, it is convenient to define
a transformed noise field
\begin{equation}
d\Omega'(x) = \mathcal{U}(x) d\Omega(x).
\end{equation}
At a given spacetime point $x$, the four real random variables
$dW'_{AR}$, $dW'_{AI}$, $dW'_{BR}$, and $dW'_{BI}$ associated with $d\Omega'(x)$
are linear combinations of the original variables
$dW_{AR}$, $dW_{AI}$, $dW_{BR}$, and $dW_{BI}$,
\begin{equation}
\begin{split}
\begin{pmatrix}
dW'_{AR} \\
dW'_{AI} \\
dW'_{BR} \\
dW'_{BI}
\end{pmatrix}
=
\tilde{\mathcal{U}}
\begin{pmatrix}
dW_{AR} \\
dW_{AI} \\
dW_{BR} \\
dW_{BI}
\end{pmatrix},
\end{split}
\end{equation}
where the mixing matrix $\tilde{\mathcal{U}}$ is orthogonal. Since
$dW_{AR}$, $dW_{AI}$, $dW_{BR}$, and $dW_{BI}$ are defined to be four independent,
identically distributed Gaussian random variables, any orthogonal transformation
of them yields another set of independent Gaussian random variables with exactly
the same probability distribution. Consequently, $d\Omega'(x)$ is statistically
equivalent to $d\Omega(x)$.

Because $H'(x)^\dag d\Omega'(x) = H(x)^\dag d\Omega(x)$ holds for any specific
realization of the noise field, we obtain the relation
\begin{equation}
\label{eq:action:su}
H'^\dag(x)\, d\Omega(x) \stackrel{d}{=} H^\dag(x)\, d\Omega(x),
\end{equation}
where $\stackrel{d}{=}$ denotes equality in probability distribution. Using
Eq.~\eqref{eq:action:su}, it follows immediately that the probability
distribution of the action~\eqref{eq:action} remains invariant under
$\mathcal{U}(x)$. The model therefore respects a statistical
$\mathrm{U}(1) \times \mathrm{SU}(2)$ symmetry.

It should be emphasized that this statistical invariance differs
fundamentally from the deterministic local gauge invariance of the
Standard Model. For a fixed realization of the stochastic source
$d\Omega(x)$, the transformed field
$\mathcal{U}(x)d\Omega(x)$ corresponds to a different realization of the
noise. Consequently, the action associated with an individual
realization is not invariant under local
$\mathrm{U}(1)\times\mathrm{SU}(2)$ transformations.
The invariance considered here refers only to the probability
distribution of the stochastic source, which remains unchanged under
such transformations. As a result, the probability distribution of
stochastic field trajectories, and hence ensemble-averaged observables,
is independent of the choice of gauge.

The present model should therefore be regarded as a stochastic
scalar-doublet toy model possessing statistical
$\mathrm{U}(1)\times\mathrm{SU}(2)$ invariance, rather than as a complete
gauge theory. As will be shown below, the stochastic dynamics generates
a nonzero ensemble average of the scalar-field magnitude,
$\langle |H| \rangle$, while the probability distribution of the
stochastic trajectories remains statistically invariant. Whether such a
statistical symmetry can be incorporated into a fully gauge-consistent
electroweak theory remains an open question for future investigation.

\section{Euler-Lagrangian equation and its solution}
\label{sec:EL}

The action~\eqref{eq:action} is formulated within the electroweak framework and
contains the full gauge structure through the covariant derivative $D_\mu$.
As discussed in Sec.~\ref{sec:model}, although the action corresponding to an
individual realization of the stochastic source is not invariant under local
$\mathrm{U}(1)\times\mathrm{SU}(2)$ transformations, its probability
distribution is invariant. Consequently, the stochastic
Euler-Lagrange equations inherit the same statistical symmetry, so that
their ensemble predictions remain invariant under local
$\mathrm{U}(1)\times\mathrm{SU}(2)$ transformations.

A complete treatment of the model would require solving the coupled
stochastic dynamics of the scalar-doublet field together with the
electroweak gauge fields. Such an analysis is beyond the scope of the
present work. Instead, we restrict attention to the gauge-field-free
sector by setting $W_\mu^a=B_\mu=0$, which allows the stochastic
dynamics of the scalar-doublet field to be studied analytically. This
restriction is introduced solely to obtain an analytically tractable
model and should not be interpreted as part of the fundamental
definition of the stochastic scalar-doublet theory. Consequently, the
results obtained in this paper should be interpreted as describing the
stochastic dynamics of the scalar-doublet field within this restricted
sector, rather than as a complete treatment of the electroweak theory.

The analysis presented below therefore focuses on the stochastic dynamics of
the scalar-doublet field within the gauge-field-free sector. Our objective is
to investigate the mathematical properties of the resulting stochastic
Euler-Lagrange equation, in particular whether its solutions develop a
nonzero ensemble average of the scalar-field magnitude and whether they
naturally generate an emergent colored-noise component. The extent to which
these features persist in the fully coupled scalar-doublet-gauge-field system
remains an open question and is left for future investigation.

Under the restriction $W_\mu^a=B_\mu=0$, the gauge-covariant derivative reduces
to the ordinary derivative, $D_\mu \rightarrow \partial_\mu$.
We then consider the classical equation of motion for the scalar-doublet field obtained
from the variational principle $\delta S=0$. In this procedure, the spacetime
metric $g_{\mu\nu}$ is treated as a fixed background field, while the stochastic
noise field $d\Omega$ is regarded as an external random source. Consequently,
the scalar-doublet field $H(x)$ is the only dynamical variable. The resulting Euler-Lagrange equation is
\begin{equation}
\label{eq:EL}
\left(
-g^{\mu\nu}\nabla_\mu\nabla_\nu
+\eta^2
\right)H
=
\gamma\,(-g)^{-1/4}\,
\frac{d\Omega}{d^4x},
\end{equation}
where $\nabla_\mu$ denotes the covariant derivative compatible with the metric
$g_{\mu\nu}$. Equation~\eqref{eq:EL} constitutes the starting point for the
subsequent analysis of stochastic $H$-field dynamics.

Equation~\eqref{eq:EL} represents four independent stochastic wave equations for
the real field components $H_{AR}$, $H_{AI}$, $H_{BR}$, and $H_{BI}$. In each case,
the left-hand side describes a massive Klein-Gordon-type operator, while the
right-hand side acts as a random-valued source. The form of
Eq.~\eqref{eq:EL} is statistically invariant under general coordinate transformations and under
homogeneous $\mathrm{U}(1) \times \mathrm{SU}(2)$ transformations, reflecting the
corresponding statistical symmetries of the action~\eqref{eq:action}. However, the local
$\mathrm{U}(1) \times \mathrm{SU}(2)$ gauge symmetry is broken at this
level due to the neglect of the gauge fields.

In the special case of flat Minkowski spacetime and a vanishing mass parameter
$\eta = 0$, Eq.~\eqref{eq:EL} reduces to the d'Alembert equation with a
random-valued source, which was studied in our previous work~\cite{Wang24}. The present equation
may therefore be viewed as a natural generalization to curved spacetime and a
nonzero mass term.

The spacetime white-noise field $d\Omega(x)$ appearing in Eq.~\eqref{eq:EL}
should be understood as a generalized stochastic process rather than an ordinary
function. The mathematical construction of stochastic differential equations of the
form of Eq.~\eqref{eq:EL}, including the definition of the continuum limit,
has been discussed in Refs.~[\onlinecite{Wang22,Wang24}].
Briefly, spacetime is first partitioned into cells of four-volume
$\Delta^4x$, and the noise field is represented by independent Gaussian random
variables associated with each cell, with variances proportional to
$\Delta^4x$. Within this discretized setting, stochastic differential equations
reduce to ordinary difference equations and stochastic integrals become finite
sums. The continuum theory is then defined by taking the limit
$\Delta^4x \rightarrow 0$. As shown in Refs.~[\onlinecite{Wang22,Wang24}],
this limit exists and is independent of the particular
choice of spacetime partition, thereby providing a well-defined mathematical
framework for the stochastic evolution equations employed throughout this work.

Wave equations of the form~\eqref{eq:EL} with deterministic sources were
systematically investigated in the classical book by
Friedlander~\cite{Friedlander}. Importantly, replacing the source term by a
random-valued function does not alter the structure of the solution method. The
key step is the construction of an appropriate Green’s function. In what follows,
we choose the retarded Green's function $G^{R}(x,y)$, defined by
\begin{equation}
\label{eq:EL:G}
\left( - g^{\mu\nu}(x) \nabla_\mu \nabla_\nu + \eta^2 \right) G^{R}(x,y)
= \frac{\delta^{(4)}(x-y)}{\sqrt{-g(x)}},
\end{equation}
where the covariant derivatives act on the coordinate $x$.
Once the retarded Green's function is known, the solution to the stochastic
equation~\eqref{eq:EL} can be written formally as
\begin{equation}
\label{Eq:EL:He}
H(x)
= \gamma \int \bigl(-g(y)\bigr)^{\frac{1}{4}} \, G^{R}(x,y)\, d\Omega(y),
\end{equation}
where $\int d\Omega(y)$ denotes the stochastic integral over spacetime.

Next, we outline the construction of the retarded Green’s function $G^{R}(x,y)$.
This construction relies on the following important theorem: if the points $x$
and $y$ lie within a geodesically convex spacetime region---namely, a region in
which any two points are connected by a unique geodesic---then there exists a
unique retarded Green's function that solves Eq.~\eqref{eq:EL:G}. The explicit
form of this solution is conveniently expressed in terms of the Synge world
function $\sigma(x,y)$, defined as one half of the squared geodesic distance
between $x$ and $y$. In a geodesically convex spacetime, $\sigma(x,y)$ is a
single-valued function.
With our sign convention, $\sigma(x,y) = 0$ corresponds to $x$ lying on a null
geodesic emanating from $y$, while $\sigma(x,y) > 0$ corresponds to $x$ lying on
a timelike geodesic from $y$. In flat Minkowski spacetime with metric
$\eta_{\mu\nu} = \mathrm{diag}(-1,1,1,1)$, the Synge function reduces to
\begin{equation}
\sigma(x,y)
= \frac{1}{2} \left[ (x^0 - y^0)^2 - \lvert \mathbf{x} - \mathbf{y} \rvert^2 \right],
\end{equation}
where $x^0$ and $\mathbf{x}$ denote the temporal and spatial coordinates of $x$,
respectively.

In the following, we consider the expanding Universe described by the
Friedmann-Lema\^{i}tre-Robertson-Walker (FLRW) metric,
\begin{equation}
ds^2 = a^2(x^0) \left[
-(dx^0)^2
+(dx^1)^2
+(dx^2)^2
+(dx^3)^2
\right],
\end{equation}
where $a(x^0)$ is the scale factor, $x^0$ denotes the conformal time, and
$\mathbf{x}=(x^1,x^2,x^3)$ represents the spatial coordinates.
We choose the Big Bang to correspond to $x^0=0$ and denote the conformal
time at the current epoch by $t_c$. We further normalize the scale factor
such that $a(t_c)=1$. With this convention, the metric tensor reduces to
$\eta_{\mu\nu}$ at the present epoch, which is convenient for discussing
the evolution of fermionic states. The explicit form of $a(x^0)$ depends
on the cosmological era under consideration. As a simplified model of a
matter-dominated universe, we adopt $a(x^0)=\frac{(x^0)^2}{t_c^2}$.

For two spacetime points $x$ and $y$ connected by a timelike or null
geodesic in an FLRW spacetime, the Synge world function can be written as
\begin{equation}\label{eq:EL:SynFLRW}
\sigma(x,y)
=
\frac{1}{2}K
\left(
\int_{y^0}^{x^0}
d\tau\,
\frac{a^2(\tau)}
{\sqrt{1+K a^2(\tau)}}
\right)^2,
\end{equation}
where the parameter $K$ is determined implicitly by
\begin{equation}
\left| \mathbf{x}-\mathbf{y} \right|
=
\int_{y^0}^{x^0}
d\tau\,
\frac{1}
{\sqrt{1+K a^2(\tau)}}.
\end{equation}
These expressions reduce to the familiar Minkowski-spacetime results when $a \equiv 1$.

Using $\sigma(x,y)$, the solution of Eq.~\eqref{eq:EL:G} can be written in the
Hadamard form
\begin{equation}
\label{eq:EL:Gs}
G^{R}(x,y)
= \frac{1}{4\pi}
\left[
U(x,y)\, \delta_{+}\!\left( \sigma \right)
+ V(x,y)\, \theta_{+}\!\left( \sigma \right)
\right],
\end{equation}
where $\delta_{+}$ and $\theta_{+}$ denote the Dirac delta function and the
Heaviside step function, respectively, with support restricted to the spacetime
region in which $x$ can be reached from $y$ by future-directed null or timelike
geodesics.

\begin{figure}[tbp]
\vspace{0.2cm}
%\vspace{1mm}.
\includegraphics[width=0.9\linewidth]{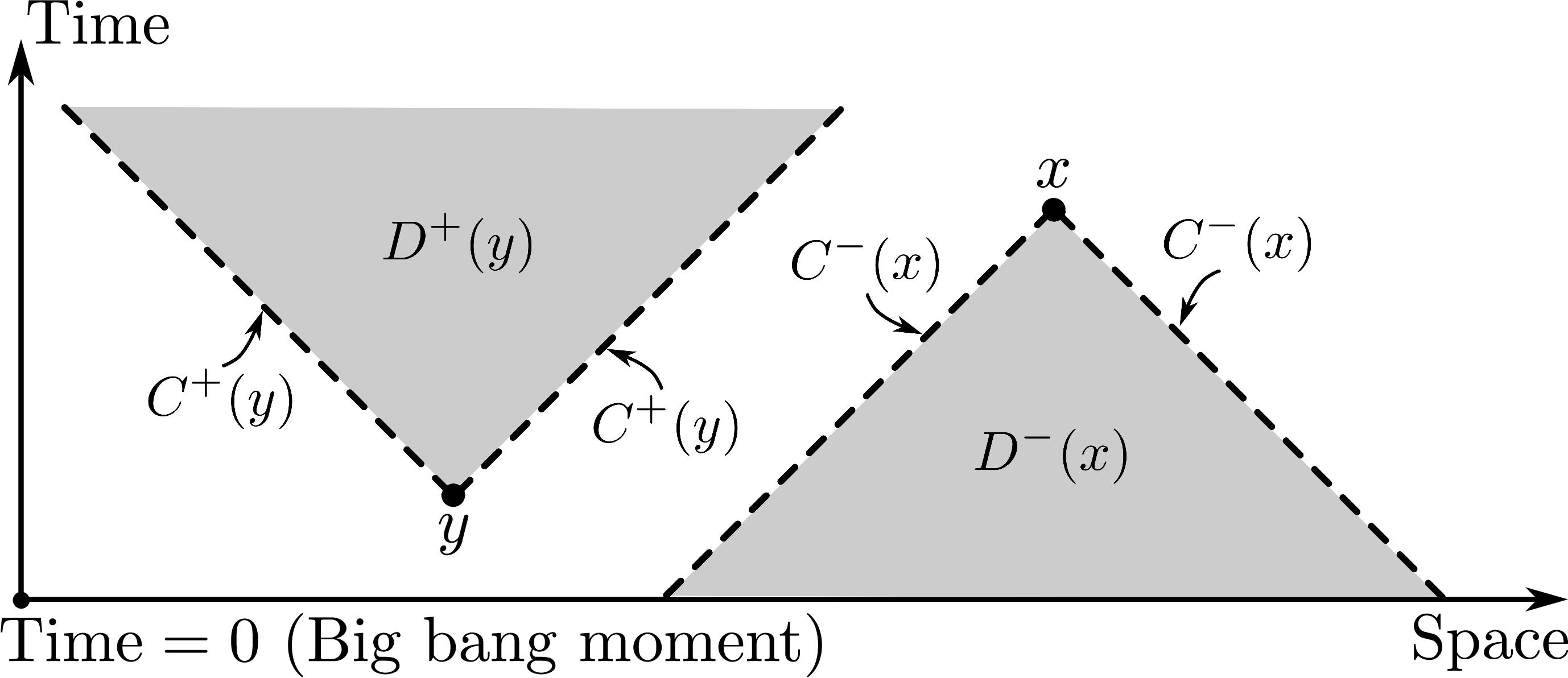}
\caption{Schematic illustration of the future null semi-cone \(C^{+}(y)\) and the future timelike domain \(D^{+}(y)\) associated with a spacetime point \(y\), as well as the past null semi-cone \(C^{-}(x)\) and the past timelike domain \(D^{-}(x)\) associated with a spacetime point \(x\), in FLRW spacetime. A null geodesic connects \(x\) and \(y\) if \(\lvert x^{0}-y^{0}\rvert=\lvert\mathbf{x}-\mathbf{y}\rvert\), whereas a timelike geodesic exists if \(\lvert x^{0}-y^{0}\rvert>\lvert\mathbf{x}-\mathbf{y}\rvert\).}\label{fig:spacetime}
\end{figure}

Let $C^{+}(y)$ and $D^{+}(y)$ denote the future null cone and the future timelike
domain of the point $y$, respectively. Then $G^{R}(x,y)$ is nonvanishing only for
$x \in C^{+}(y) \cup D^{+}(y)$. A schematic illustration of the regions $C^{+}(y)$
and $D^{+}(y)$ is shown in Fig.~\ref{fig:spacetime}. The first term in
Eq.~\eqref{eq:EL:Gs}, known as the \emph{$\delta$-term}, contributes only when
$x \in C^{+}(y)$, or equivalently when $\sigma(x,y) = 0$ and $x^{0} \ge y^{0}$.
The second term, commonly referred to as the \emph{tail term}, contributes only
when $x \in D^{+}(y)$, corresponding to $\sigma(x,y) > 0$ and $x^{0} \ge y^{0}$.

It is important to note that the Hadamard representation of the
retarded Green's function in Eq.~\eqref{eq:EL:Gs} is strictly valid
when the spacetime points $x$ and $y$ lie within a common
geodesically convex neighborhood, ensuring that Synge's world
function $\sigma(x,y)$ is uniquely defined.
Flat Minkowski spacetime satisfies this condition globally.
Generic FLRW spacetimes are not globally geodesically convex in the
strict mathematical sense. Nevertheless, geodesically convex normal
neighborhoods exist locally, within which the Hadamard construction
remains valid. In the present work, the geodesic configurations
relevant to the evaluation of the Green's function admit unique
connecting geodesics and lie within such neighborhoods. Consequently,
the Synge function $\sigma(x,y)$ and the corresponding retarded
Green's function are well defined throughout our analysis.

We now explain how to determine the smooth functions $U(x,y)$ and $V(x,y)$
appearing in Eq.~\eqref{eq:EL:Gs}. Their explicit forms depend on the background
metric $g_{\mu\nu}$, and closed-form analytical expressions can be obtained only
in a limited number of cases, such as flat Minkowski spacetime
($g_{\mu\nu} = \eta_{\mu\nu}$) and FLRW spacetime with
$a(x^0) = \left. {(x^0)^2} \middle/ {t_c^2} \right.$.

For a fixed point $y$ and a point $x \in C^{+}(y)$ (i.e., $x$ lies on the future
null geodesic of $y$), the function $U(x,y)$ satisfies the transport equation
\begin{equation}
\label{eq:EL:eU}
\left( g^{\mu\nu} \nabla_\mu \nabla_\nu \sigma \right) U
+ 2 g^{\mu\nu} \partial_\mu \sigma \, \partial_\nu U
+ 4 U = 0,
\end{equation}
together with the initial condition $U(y,y) = 1$. By connecting $x$ and $y$ with
their unique null geodesic and integrating Eq.~\eqref{eq:EL:eU} along this
geodesic from $y$ to $x$, one obtains a unique solution,
\begin{equation}
\left. U(x,y) \right|_{x \in C^{+}(y)}
= \frac{ \sqrt{ \left| \det S \right| } }
{ \left| g(x) g(y) \right|^{1/4} },
\end{equation}
where $g(x)$ and $g(y)$ denote the determinants of the metric tensor
$g_{\mu\nu}$ evaluated at $x$ and $y$, respectively. The $4 \times 4$ matrix $S$
is defined by its components $S_{\mu\nu}
= \left. {\partial^2 \sigma(x,y)} \middle/ {\partial x^\mu \, \partial y^\nu} \right.$.
For flat spacetime and for FLRW spacetime with
$a(x^0) = (x^0)^2 / t_c^2$, the function $U(x,y)$ takes the explicit form
\begin{equation}
\label{eq:EL:Ua}
\left. U(x,y) \right|_{x \in C^{+}(y)}
=
\begin{cases}
1, & \text{flat spacetime}, \\[1.2ex]
\displaystyle
\frac{\int_{y^0}^{x^0} d\tau \, a^2(\tau)}
{ \left| \mathbf{x} - \mathbf{y} \right| \,
a(x^0)\, a(y^0) },
& \text{FLRW spacetime}.
\end{cases}
\end{equation}
It is clear from Eq.~\eqref{eq:EL:Gs} that knowledge of $U(x,y)$ for
$x \in C^{+}(y)$ is sufficient for constructing the retarded Green's function,
since the factor $\delta_{+}(\sigma)$ multiplying $U(x,y)$ eliminates any
contribution from points $x \notin C^{+}(y)$.

Having obtained $U(x,y)$, we now proceed to construct the function $V(x,y)$.
The determination of $V(x,y)$ involves two successive steps. The first step is to
solve the transport equation
\begin{equation}
\label{eq:EL:V1}
\left( g^{\mu\nu} \nabla_\mu \nabla_\nu \sigma \right) V
+ 2 g^{\mu\nu} \partial_\mu \sigma \, \partial_\nu V
+ 2 V
+ g^{\mu\nu} \nabla_\mu \nabla_\nu U
= 0,
\end{equation}
which is valid only on the future null cone, $x \in C^{+}(y)$, where
$\sigma(x,y) = 0$. For a fixed point $y$, Eq.~\eqref{eq:EL:V1} can again be solved by
connecting $x$ to $y$ with their unique null geodesic and integrating along this
geodesic. This procedure determines the value of $V(x,y)$ on the null cone.
For flat spacetime and for FLRW spacetime with
$a(x^0) = (x^0)^2 / t_c^2$, we obtain
\begin{equation}
\label{eq:EL:V0th}
\left. V(x,y) \right|_{x \in C^{+}(y)}
=
\begin{cases}
-\dfrac{\eta^2}{2},
& \text{flat}, \\[1.2ex]
\displaystyle
\frac{t_c^4}{(x^0)^3 (y^0)^3}
-
\frac{\eta^2}{10}
\frac{(x^0)^5 - (y^0)^5}
{(x^0 - y^0)(x^0)^2 (y^0)^2},
& \text{FLRW}.
\end{cases}
\end{equation}

The second step is to determine $V(x,y)$ for points
$x \in D^{+}(y)$, where $\sigma(x,y) > 0$. In this region, $V(x,y)$ satisfies the
homogeneous wave equation
\begin{equation}
\label{eq:EL:V2}
g^{\mu\nu} \nabla_\mu \nabla_\nu V = 0.
\end{equation}
A systematic approach to solving Eq.~\eqref{eq:EL:V2} is to expand $V(x,y)$ as a power
series in $\sigma(x,y)$ and to determine the coefficients order by order, with
Eq.~\eqref{eq:EL:V0th} serving as the zeroth-order term.
Starting from the FLRW solution in Eq.~\eqref{eq:EL:V0th}, however, leads to expressions
that are analytically intractable. Fortunately, the conformal time $t_c$ at the
current epoch is cosmologically large, and the spacetime geometry near the present
epoch is approximately flat on large scales. Consequently, when both $x^0$ and
$y^0$ are close to $t_c$, the FLRW solution is well approximated by the flat-spacetime
result $V \simeq -\eta^2/2$. Moreover, for a rough estimation of the field $H(x)$ at
the current epoch, only values of $V(x,y)$ with $x^0 \approx y^0 \approx t_c$ are
required. We therefore adopt
$\left. V(x,y) \right|_{x \in C^{+}(y)} \approx -\left. {\eta^2} \middle/ {2} \right.$
as the starting point of the power-series expansion, and simultaneously approximate
the differential operator in Eq.~\eqref{eq:EL:V2} by its flat-spacetime form,
$g^{\mu\nu} \nabla_\mu \nabla_\nu \to \eta^{\mu\nu} \partial_\mu \partial_\nu$.
Under these approximations, the resulting power series converges to
\begin{equation}
\label{eq:EL:Va}
V(x,y)
=
-\frac{\eta^2}{2}
\sum_{n=0}^{\infty}
\frac{\left( - \eta^2 \sigma \right)^n}
{2^n \, n! \, (n+1)!}
=
-\eta^2
\frac{J_1\!\left( \sqrt{2 \eta^2 \sigma} \right)}
{\sqrt{2 \eta^2 \sigma}},
\end{equation}
where $J_1$ denotes the Bessel function of the first kind.

The expression~\eqref{eq:EL:Va} should be regarded as an analytical approximation
to the exact FLRW solution of Eq.~\eqref{eq:EL:V2}. Its purpose is to provide
explicit analytical estimates for the stochastic scalar-doublet field at the
present cosmological epoch. A quantitatively more accurate treatment would
require solving Eq.~\eqref{eq:EL:V2} directly in the FLRW background. Such a calculation
lies beyond the scope of the present work and is left for future investigation.

Substituting the expressions for $U(x,y)$ and $V(x,y)$ from
Eqs.~\eqref{eq:EL:Ua} and~\eqref{eq:EL:Va} into Eq.~\eqref{eq:EL:Gs}, we obtain the
retarded Green's function $G^R(x,y)$. Inserting $G^R(x,y)$ into
Eq.~\eqref{Eq:EL:He} then yields an explicit solution to the
Euler-Lagrange equation for the field $H(x)$.
Since the retarded Green's function consists of a $\delta$-term and a tail term,
the scalar-doublet field can be correspondingly decomposed as
\begin{equation}
\label{eq:EL:HdHt}
H(x) = H^{(\delta)}(x) + H^{(\theta)}(x),
\end{equation}
with
\begin{equation}
\label{eq:EL:2He}
\begin{aligned}
H^{(\delta)}(x)
&= \frac{\gamma}{4\pi}
\int d\Omega(y)\,
\left( -g(y) \right)^{1/4}
U(x,y)\,
\delta_{+}\!\left( \sigma(x,y) \right), \\
H^{(\theta)}(x)
&= \frac{\gamma}{4\pi}
\int d\Omega(y)\,
\left( -g(y) \right)^{1/4}
V(x,y)\,
\theta_{+}\!\left( \sigma(x,y) \right).
\end{aligned}
\end{equation}

For a fixed spacetime point $x$, the stochastic integrals in
Eq.~\eqref{eq:EL:2He} extend over all points $y$ for which $\delta_{+}$ or
$\theta_{+}$ is nonvanishing. As discussed above, $\delta_{+}$ and $\theta_{+}$
are nonzero only when $x \in C^{+}(y)$ and $x \in D^{+}(y)$, respectively.
Equivalently, for a fixed $x$, the integration domain for $y$ is restricted to
$y \in C^{-}(x)$ and $y \in D^{-}(x)$, i.e., points that can be reached from $x$
by past-directed null and timelike geodesics, respectively. These regions are
schematically illustrated in Fig.~\ref{fig:spacetime}.

Since the sets $C^{-}(x)$ and $D^{-}(x)$ do not intersect, and 
$d\Omega(y)$ at different $y$ are independent random variables, the two contributions
$H^{(\delta)}(x) \sim \int_{y \in C^{-}(x)} d\Omega(y)$ and
$H^{(\theta)}(x) \sim \int_{y \in D^{-}(x)} d\Omega(y)$ are statistically
independent. The scalar-doublet field therefore naturally decomposes into two
independent stochastic components. As will be shown in the following sections,
these two components possess distinct statistical properties and give rise to
different physical effects when coupled to fermions through a Yukawa-type
interaction.

The first equation of Eq.~\eqref{eq:EL:2He} defining $H^{(\delta)}(x)$ contains the distribution
$\delta_{+}(\sigma(x,y))$, whose support is restricted to the
past light cone of $x$. Consequently, the expression cannot be
interpreted as an ordinary four-dimensional stochastic integral with a
pointwise integrand. Instead, following the standard treatment of
distribution-valued fields, we define $H^{(\delta)}(x)$ as a random distribution through its action on
arbitrary smooth compactly supported test functions $f(x)$.

More precisely, the pairing between
$H^{(\delta)}$ and a test function $f$ is defined by
\begin{equation}
\begin{split}
& \langle H^{(\delta)},f\rangle
\equiv 
\int d^4x\,H^{(\delta)}(x)f(x)
\\ & =
\frac{\gamma}{4\pi}
\int d^4x\,f(x)
\int d\Omega(y)
(-g(y))^{1/4}
U(x,y)
\delta_{+}(\sigma(x,y)).
\end{split}
\end{equation}
Interchanging the order of integration gives
\begin{equation}
\begin{split}
\langle H^{(\delta)},f\rangle
=
\frac{\gamma}{4\pi}
\int
d\Omega(y)
(-g(y))^{1/4}
F(y),
\end{split}
\end{equation}
where $F(y) =
\int d^4x\,
f(x) U(x,y)
\delta_{+}(\sigma(x,y))$.
Since $f(x)$ and $U(x,y)$ are ordinary functions, the above pairing with the distribution
$\delta_{+}(\sigma)$ is well defined, so that
$F(y)$ is an ordinary function of $y$.
The remaining stochastic integral
$\int d\Omega(y)(-g(y))^{1/4}F(y)$
is precisely the spacetime stochastic integral defined above through
the continuum limit of stochastic sums.
This construction therefore defines
$H^{(\delta)}$ as a well-defined distribution-valued stochastic field.

\section{Statistical Symmetry Breaking and Ensemble Average of the Scalar-Doublet Magnitude}
\label{sec:Htt}

We now turn to the properties of $H^{(\theta)}(x)$, namely the tail contribution
to the scalar-doublet field. As is evident from the second line of
Eq.~\eqref{eq:EL:2He}, $H^{(\theta)}(x)$ consists of four components,
$H^{(\theta)}_{AR}(x)$, $H^{(\theta)}_{AI}(x)$, $H^{(\theta)}_{BR}(x)$, and
$H^{(\theta)}_{BI}(x)$. Since the underlying noise variables
$dW_{AR}$, $dW_{AI}$, $dW_{BR}$, and $dW_{BI}$ are independent and identically
distributed Gaussian random numbers, the four components of the scalar-doublet field are
likewise independent and identically distributed. Consequently, it suffices to
analyze a single component, as the properties of the remaining three follow
immediately.

Without loss of generality, we focus on the component $H^{(\theta)}_{BR}(x)$.
Using the approximate form of $V(x,y)$ obtained in
Eq.~\eqref{eq:EL:Va}, it can be written as
\begin{equation}
\label{eq:m:HBR}
H^{(\theta)}_{BR}(x)
\approx
-\frac{\eta^2 \gamma}{4\pi}
\int_{y \in D^{-}(x)} dW_{BR}(y)\,
\frac{J_1\!\left( \sqrt{2\eta^2\sigma(x,y)} \right)}
{\sqrt{2\eta^2\sigma(x,y)}},
\end{equation}
where we have further used the approximation $g(y) \approx 1$. This approximation
is valid when $y^0$ is close to the present conformal time $t_c$, which is the same
condition adopted in the construction of $V(x,y)$. As a result,
Eq.~\eqref{eq:m:HBR} provides a reliable description of $H^{(\theta)}_{BR}(x)$ only
when both $x^0$ and $y^0$ are close to $t_c$.

We emphasize that the approximation $g(y)\approx 1$ and the use of the
flat-spacetime expression for $V(x,y)$ are introduced solely to obtain
analytical estimates of $H^{(\theta)}(x)$ appearing below.
These approximations are expected to be reliable for spacetime regions
sufficiently close to the present cosmological epoch, where the FLRW
metric deviates only weakly from flat spacetime. However,
contributions originating from very early cosmological times may
receive non-negligible corrections due to the evolution of the scale
factor and the exact curved-spacetime form of the tail function
$V(x,y)$. A quantitatively accurate treatment would therefore require
retaining the full FLRW expression for $V(x,y)$ over the entire
integration domain, which would most likely necessitate numerical
methods. Such an analysis lies beyond the scope of the present work
and is left for future investigation.

It is important to emphasize, however, that the qualitative properties of
$H^{(\theta)}_{BR}(x)$ do not depend on the specific approximations used here.
From the exact expression in Eq.~\eqref{eq:EL:2He}, $H^{(\theta)}_{BR}(x)$ is seen
to be a weighted sum of independent Gaussian random variables $dW_{BR}(y)$, with
$\left(-g(y)\right)^{1/4} V(x,y)$ acting as the weight. Employing more accurate
expressions for $g(y)$ and $V(x,y)$ would refine the quantitative estimates but
would not alter the qualitative statistical behavior of the field.

Since $H^{(\theta)}_{BR}(x)$ is a sum of independent Gaussian random variables,
it must itself be Gaussian distributed. A Gaussian random variable is completely
characterized by its mean and variance. The mean value of $H^{(\theta)}_{BR}(x)$
vanishes identically, because
$\langle dW_{BR}(y) \rangle \equiv 0$, where $\langle \cdot \rangle$ denotes the
expectation value over realizations of the noise field. The variance,
\begin{equation}
D(x) \equiv \left\langle \left( H^{(\theta)}_{BR}(x) \right)^2 \right\rangle,
\end{equation}
can be evaluated directly from Eq.~\eqref{eq:m:HBR}, yielding
\begin{equation}
\label{eq:m:Dd}
D(x)
=
\frac{\gamma^2 \eta^4}{16\pi^2}
\int_{y \in D^{-}(x)} d^4 y\,
\left[
\frac{J_1\!\left( \sqrt{2\eta^2\sigma(x,y)} \right)}
{\sqrt{2\eta^2\sigma(x,y)}}
\right]^2.
\end{equation}
This expression involves a four-dimensional integral over the past timelike
domain $D^{-}(x)$.

At this point, it becomes clear why an expanding spacetime with a finite initial
time is essential for the consistency of our theory. In flat Minkowski spacetime,
the past timelike domain $D^{-}(x)$ extends infinitely far into the past, causing
the integral $\int_{y \in D^{-}(x)} d^4 y$ to diverge. This infrared divergence in
the variance renders $H^{(\theta)}_{BR}(x)$ ill defined as a random variable.
By contrast, in an FLRW spacetime with a finite initial time, where $x^0 \geq 0$,
and $x^0 = 0$ corresponds to the Big Bang, the domain $D^{-}(x)$ has finite
four-volume. As a result, the integral in Eq.~\eqref{eq:m:Dd} converges (see
Fig.~\ref{fig:spacetime} for a schematic illustration of $D^{-}(x)$ in FLRW
spacetime), and $H^{(\theta)}_{BR}(x)$ becomes a well-defined Gaussian random
variable.

We emphasize that the absence of infrared divergence does not rely on
the approximations employed in Eq.~\eqref{eq:m:HBR}. The essential ingredient is the existence of a
finite initial time in FLRW cosmology, which restricts the integration
domain to the finite past causal domain of the spacetime point under
consideration. It is this finite integration region that provides the
underlying reason for the convergence of the stochastic integrals.
Consequently, we expect the infrared convergence to persist when the
exact FLRW expressions for $g(y)$ and $V(x,y)$ are employed. While
these more accurate expressions may modify the quantitative estimates,
they are not expected to alter the qualitative conclusion that
$H^{(\theta)}_{BR}(x)$ remains a well-defined Gaussian random variable.

For a generic FLRW metric, the Synge function $\sigma(x,y)$ has a complicated
form (see Eq.~\eqref{eq:EL:SynFLRW}), which renders the integral in Eq.~\eqref{eq:m:Dd} difficult to evaluate
analytically. To obtain a tractable expression, we adopt a further approximation
and replace $\sigma(x,y)$ by its flat-spacetime form,
$\sigma(x,y) \approx \frac{1}{2}
\left[ (x^0 - y^0)^2 - \left| \mathbf{x} - \mathbf{y} \right|^2 \right]$.
With this approximation, Eq.~\eqref{eq:m:Dd} can be reduced to
\begin{equation}
\label{eq:m:D1}
\begin{split}
D(x)
= & \ \frac{\gamma^2}{8\pi}
\int_0^{\tilde{x}^0} ds \,
\left( J_1(s) \right)^2 s
\\ & \times
\left[
\frac{\tilde{x}^0}{s}
\sqrt{\left( \frac{\tilde{x}^0}{s} \right)^2 - 1 }
-
\ln \!\left(
\frac{\tilde{x}^0}{s}
+
\sqrt{\left( \frac{\tilde{x}^0}{s} \right)^2 - 1 }
\right)
\right],
\end{split}
\end{equation}
where we have introduced the dimensionless rescaled time
$\tilde{x}^0 \equiv \eta x^0$. This quantity is dimensionless because
$\eta$ and $x^0$ carry dimensions $\mathrm{L}^{-1}$ and $\mathrm{L}$,
respectively.

We are primarily interested in the regime $x^0 \sim t_c$, where the conformal
time is cosmologically large. As will be shown later, a rough estimate of $\eta$
implies that $\tilde{x}^0 = \eta x^0$ is likewise a very large number.
Meanwhile, the Bessel function $J_1(s)$ is oscillatory with an amplitude that
decays as $1/\sqrt{s}$, so integrals involving $J_1(s)$ converge rapidly once
$s$ exceeds a few tens. Therefore, to leading order, we may extend the upper
limit of the integral in Eq.~\eqref{eq:m:D1} to $+\infty$ and simultaneously
simplify the bracketed term by using the asymptotic condition $\tilde{x}^0 / s \gg 1$.
Under these approximations, the leading contribution to $D(x)$ becomes
\begin{equation}
\label{eq:m:Dr}
\begin{split}
D(x) \approx & \ \frac{\gamma^2 \left(\tilde{x}^0\right)^2}{8\pi}
\int_0^{\infty} ds \,
\left( J_1(s) \right)^2 s
\\ & \times
\left[
\frac{1}{s}
\sqrt{\frac{1}{s^2} - \frac{1}{\left(\tilde{x}^0\right)^2}  }
- \frac{1}{\left(\tilde{x}^0\right)^2} 
\ln \!\left(
\frac{\tilde{x}^0}{s}
+
\sqrt{\left( \frac{\tilde{x}^0}{s} \right)^2 - 1 }
\right)
\right] \\
\approx & \ \frac{\gamma^2 \eta^2 (x^0)^2}{8\pi}
\int_0^{\infty} ds \,
\frac{\left( J_1(s) \right)^2}{s}
\\[4pt]
= & \ \frac{\gamma^2 \eta^2}{16\pi} (x^0)^2 ,
\end{split}
\end{equation}
where we have neglected the logarithmic term in Eq.~\eqref{eq:m:D1}, since it is
subleading compared with the dominant power-law contribution. According to
Eq.~\eqref{eq:m:Dr}, the variance of $H^{(\theta)}_{BR}(x)$ depends only on the
conformal time $x^0$ and grows quadratically with it.

At the Big Bang moment, $x^0 = 0$, Eq.~\eqref{eq:m:Dr} immediately yields
$D(0) = 0$. From probability theory, a random variable with zero mean and zero
variance must be identically zero. Thus, the scalar-doublet field satisfies
$H(x^0 = 0) \equiv 0$. This conclusion can also be obtained directly from
Eq.~\eqref{eq:EL:2He}. Thus, the scalar-doublet field vanishes at the initial time, and the
deterministic $\mathrm{U(1)}\times\mathrm{SU(2)}$ symmetry remains unbroken at
that instant.

As the Universe evolves and the conformal time $x^0$ increases, the variance
$D(x)$ grows continuously, corresponding to an increasingly broadened
distribution of the scalar-doublet-field components. A finite variance implies that
$H(x)$ becomes a genuinely random field. 
For $x^0>0$, individual realizations of the scalar-doublet field are
generically nonvanishing and therefore do not preserve the deterministic
$\mathrm{U(1)}\times\mathrm{SU(2)}$ symmetry.

At the same time, because the four components
$\left( H^{(\theta)}_{AR}, H^{(\theta)}_{AI}, H^{(\theta)}_{BR},
H^{(\theta)}_{BI} \right)$ are independent and identically distributed Gaussian
random variables, any $\mathrm{U(1)} \times \mathrm{SU(2)}$ transformation leaves
the probability distribution of $H$ invariant. Hence, even for $x^0 > 0$, the
scalar-doublet field continues to respect a statistical
$\mathrm{U(1)} \times \mathrm{SU(2)}$ symmetry.

The evolution of the four-component scalar-doublet field therefore resembles a Brownian
motion in a four-dimensional internal space, in the sense that each component
follows an independent Gaussian distribution with zero mean. The difference,
however, is equally important: while the variance in Brownian motion grows
linearly with time, the variance of the scalar-doublet-field components in our model
grows quadratically with the conformal time $x^0$.

\subsection{Probability Distribution and Ensemble Average of the Scalar-Doublet Magnitude}
\label{sec:EH}

The quantity of primary interest in the present stochastic scalar-doublet model
is the gauge-invariant magnitude of the field. For an arbitrary scalar-doublet
configuration
$\left( H^{(\theta)}_{AR}, H^{(\theta)}_{AI}, H^{(\theta)}_{BR},
H^{(\theta)}_{BI} \right)$,
one may perform a
$\mathrm{U}(1)\times\mathrm{SU}(2)$ transformation such that
\begin{equation}
\label{eq:m:HctH}
\begin{split}
\left(
\begin{array}{c}
H^{(\theta)}_{AR}+iH^{(\theta)}_{AI}\\
H^{(\theta)}_{BR}+iH^{(\theta)}_{BI}
\end{array}
\right)
\stackrel{\mathcal U}{\longrightarrow}
\left(
\begin{array}{c}
0\\
|H|
\end{array}
\right),
\end{split}
\end{equation}
where
\begin{equation}
|H|
=
\sqrt{
\left(H^{(\theta)}_{AR}\right)^2
+
\left(H^{(\theta)}_{AI}\right)^2
+
\left(H^{(\theta)}_{BR}\right)^2
+
\left(H^{(\theta)}_{BI}\right)^2
}.
\end{equation}
The quantity $|H|$ is invariant under
$\mathrm{U}(1)\times\mathrm{SU}(2)$ transformations and therefore provides a
natural gauge-independent characterization of the stochastic scalar-doublet
field. In the following, we investigate the statistical properties of this
gauge-invariant quantity.

As established above, the four components of the scalar-doublet field are
independent and identically distributed Gaussian random variables with zero
mean and variance $D(x)$. Their joint probability density is therefore
\begin{equation}
\begin{split}
&
P\!\left(
H^{(\theta)}_{AR},
H^{(\theta)}_{AI},
H^{(\theta)}_{BR},
H^{(\theta)}_{BI}
\right)
=
\frac{1}{(2\pi D)^2}
\\
&\times
\exp\!\left[
-\frac{
\left(H^{(\theta)}_{AR}\right)^2
+
\left(H^{(\theta)}_{AI}\right)^2
+
\left(H^{(\theta)}_{BR}\right)^2
+
\left(H^{(\theta)}_{BI}\right)^2
}{2D}
\right].
\end{split}
\end{equation}
Since $|H|$ depends only on the radial distance in this four-dimensional field
space, its probability density is obtained by integrating over the angular
degrees of freedom. The resulting distribution is
\begin{equation}
\label{eq:m:PHa}
P(|H|)
=
\frac{|H|^3}{2D^2}
\exp\!\left(
-\frac{|H|^2}{2D}
\right),
\end{equation}
which is the Maxwell-type distribution associated with a four-dimensional
Gaussian random field.

\begin{figure}[tbp]
\vspace{0.2cm}
%\vspace{1mm}.
\includegraphics[width=0.8\linewidth]{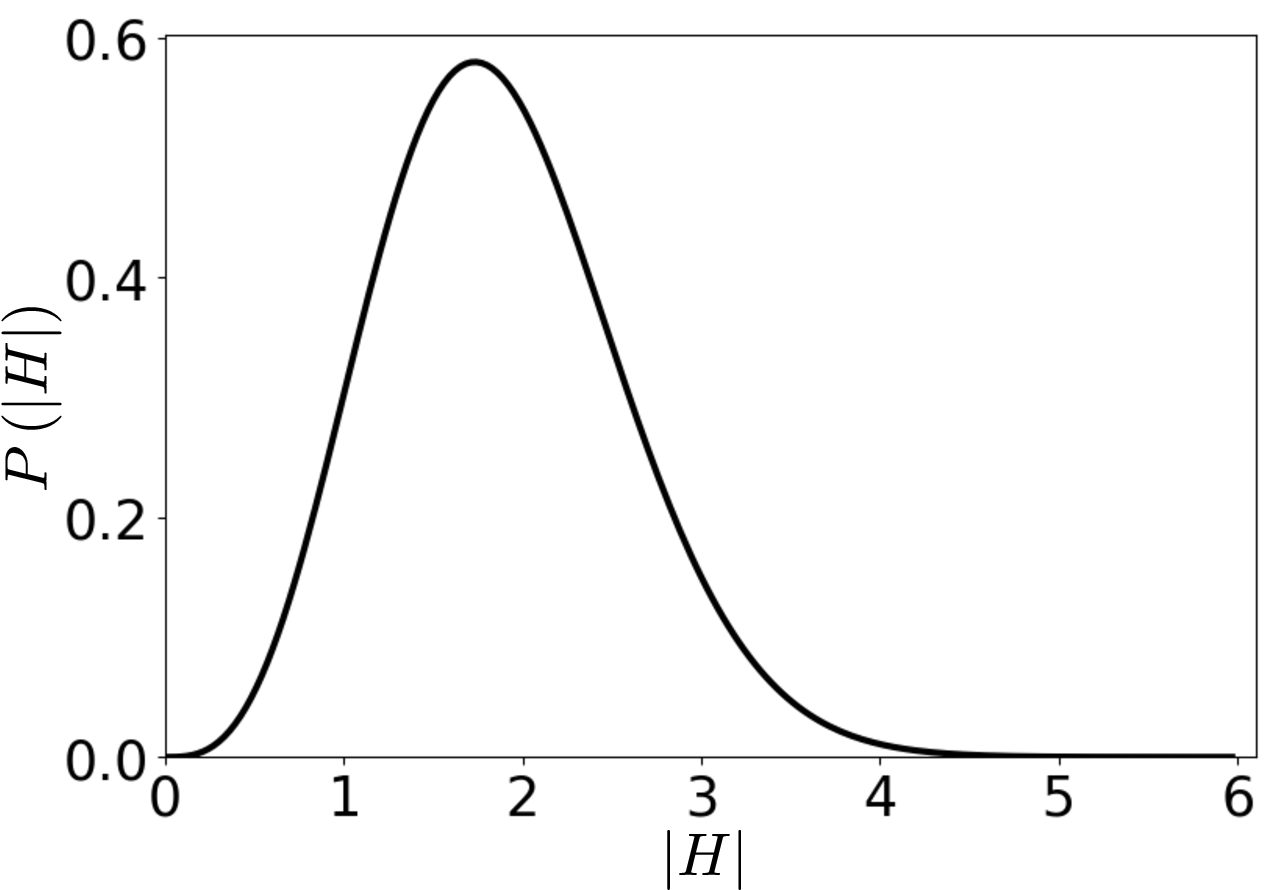}
\caption{Schematic illustration of the probability density function $P\!\left(\lvert H\rvert\right)$, shown for the representative case $D=1$.}\label{fig:PH}
\end{figure}

Figure~\ref{fig:PH} displays the shape of the probability distribution
$P\!\left( \lvert H \rvert \right)$. From this distribution, it is clear that
both the location of its maximum and its expectation value are nonzero. The
maximum occurs at $\lvert H \rvert = \sqrt{3D}$, while the expectation value is
given by
\begin{equation}
\label{eq:m:Hva}
\left\langle \lvert H \rvert \right\rangle
= \frac{3\sqrt{2\pi D}}{4}
= \frac{3 \gamma \eta}{8\sqrt{2}} \, x^0 .
\end{equation}
Although each individual component of the scalar-doublet field follows a Gaussian
distribution whose maximum is located at zero, the probability density of
$\lvert H \rvert$ peaks at a nonzero value. This behavior arises because the
scalar-doublet field has four independent components: the volume of the four-dimensional
field space contained in the interval
$\left[ \lvert H \rvert, \lvert H \rvert + d\lvert H \rvert \right]$ grows rapidly
with $\lvert H \rvert$, causing $P(\lvert H \rvert)$ to increase for small
$\lvert H \rvert$ before eventually decaying.

The maximum value $\sqrt{3D}$ and the ensemble average
$\left\langle \lvert H \rvert \right\rangle \approx 1.88 \sqrt{D}$
are numerically close. Either quantity therefore provides a representative
measure of the typical magnitude of the stochastic scalar-doublet field.
Throughout the remainder of this paper, we use the ensemble average
$\left\langle \lvert H \rvert \right\rangle$ as the characteristic field
magnitude when discussing illustrative applications of the model.
According to Eq.~\eqref{eq:m:Hva}, this characteristic magnitude grows
linearly with conformal time and is determined by the product
$\gamma \eta$. Possible choices for the parameters $\gamma$ and $\eta$,
together with their physical interpretation, are briefly discussed in
Sec.~\ref{sec:para}.

There is a fundamental difference between the stochastic scalar-doublet
dynamics considered here and the conventional scalar-field dynamics
described by the Standard Model Higgs sector. In the Standard Model,
the vacuum expectation value is determined by the minimum of the Higgs
potential, $V(\lvert H \rvert) = -\mu^2 \lvert H \rvert^2 + \lambda \lvert H \rvert^4$,
and becomes essentially constant after electroweak symmetry breaking.
By contrast, in the present toy model, the characteristic magnitude of
the scalar-doublet field arises from stochastic driving rather than from
a self-interaction potential. Within this framework, the nonzero ensemble
average of $\lvert H \rvert$ reflects the statistical properties of the
noise-driven field rather than the existence of a deterministic vacuum.

The present analysis further suggests that the ensemble average
$\langle |H| \rangle$ may evolve with conformal time. Near the present
cosmological epoch, the approximate analytical solution predicts an
approximately linear dependence on $x^0$. Whether this behavior persists
throughout the full cosmological history requires solving the stochastic
field equations using the exact FLRW Green function, which lies beyond
the scope of the present work.

Although the present model is formulated as a stochastic scalar-doublet
toy model, it is nevertheless natural to ask how such stochastic dynamics
might manifest itself if a similar mechanism were incorporated into a
more complete electroweak framework. For example, if the scalar doublet
were identified with the Higgs field and coupled to fermions through
Yukawa interactions, a time-dependent ensemble average of
$\langle |H| \rangle$ could, in principle, induce a corresponding evolution
of effective fermion mass scales. Such a possibility might eventually have
implications for cosmological observables, including the interpretation of
spectral observations. However, establishing such connections requires a
gauge-consistent electroweak formulation together with a quantitative
phenomenological analysis, neither of which is attempted in the present
work. These questions are therefore left for future investigation.

\subsection{Spatial Fluctuation of the Scalar-Doublet Field}
\label{sec:HFF}

\begin{figure}[tbp]
\vspace{0.2cm}
%\vspace{1mm}.
\includegraphics[width=0.8\linewidth]{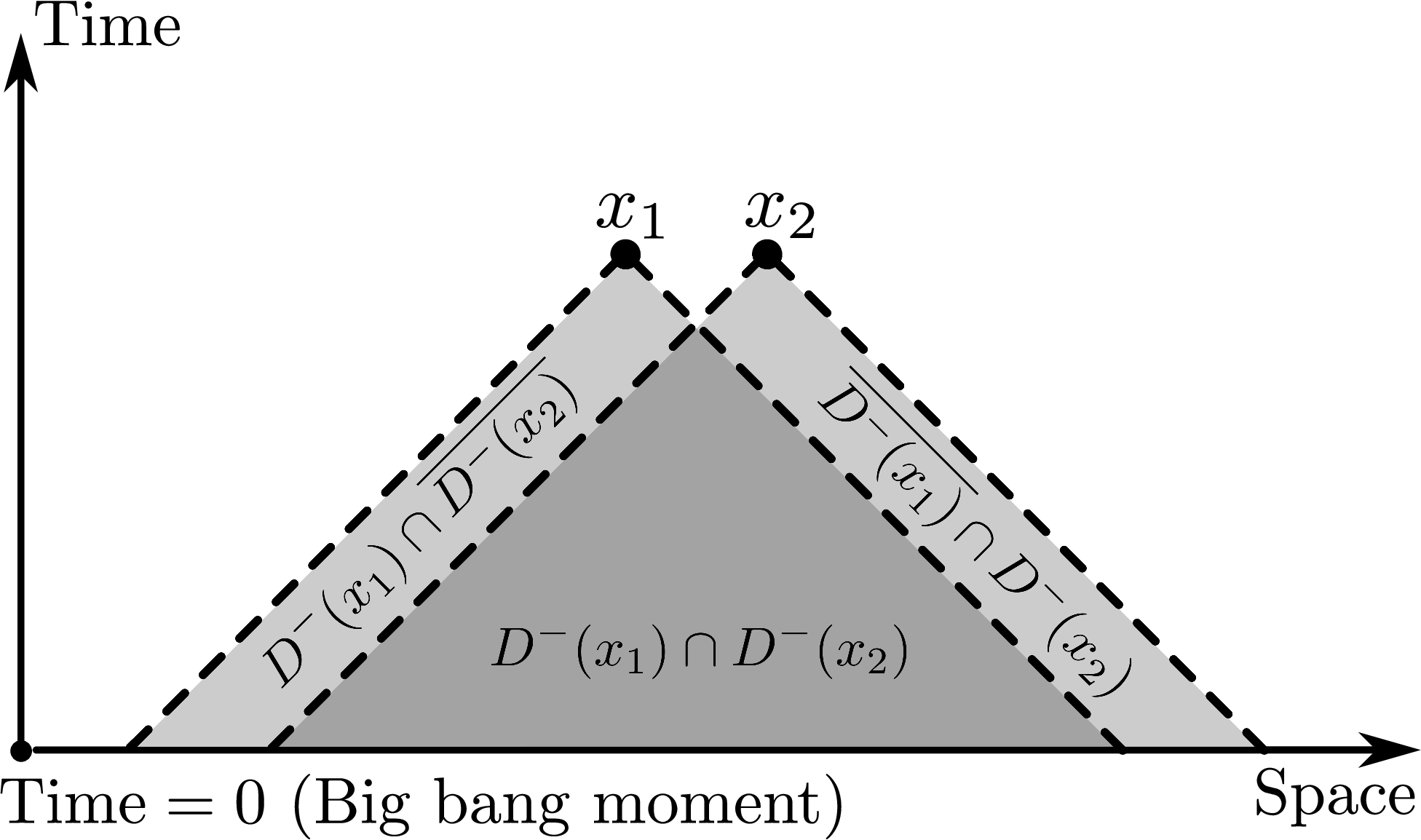}
\caption{Supporting regions of the scalar-doublet field at two spacetime points,
$x_1 = (x^0,\mathbf{x}_1)$ and $x_2 = (x^0,\mathbf{x}_2)$,
which share the same temporal coordinate but differ in spatial position.
The heavily shaded area denotes the overlap between the two supporting regions.}\label{fig:Hs}
\end{figure}

Another characteristic feature of the present stochastic scalar-doublet model
is the emergence of spatial fluctuations in the field magnitude
$\left|H \right| $. Since the scalar-doublet field is constructed from
stochastic integrals whose integration domains depend on the spacetime
point, two distinct spatial locations generally involve different subsets
of the underlying noise field. Consequently, even at a fixed conformal time, the
magnitude $|H|$ becomes a spatially fluctuating random variable with a finite
correlation length. In this subsection, we analyze the statistical properties
of these fluctuations and estimate their characteristic spatial scale.

The existence of such fluctuations is already implied by the finite width of the
probability distribution $P\!\left(\lvert H \rvert\right)$ shown in
Fig.~\ref{fig:PH}. A natural quantitative measure of the fluctuation strength is
the variance of $\lvert H \rvert$. Using Eq.~\eqref{eq:m:PHa}, we readily obtain
\begin{equation}
\label{eq:m:VaHH}
\mathrm{Var}\!\left(\lvert H \rvert\right)
= \left\langle \lvert H \rvert^2 \right\rangle
- \left\langle \lvert H \rvert \right\rangle^2
\approx 0.47 D ,
\end{equation}
which is of the same order as
$\left\langle \lvert H \rvert \right\rangle^2 \approx 3.53 D$.
The ratio between the standard deviation
$\Delta \lvert H \rvert = \sqrt{\mathrm{Var}\!\left(\lvert H \rvert\right)}$ and the expectation value
$\left\langle \lvert H \rvert \right\rangle$ is therefore a constant,
approximately $0.36$. Since
$\left\langle \lvert H \rvert \right\rangle$ increases with time, the absolute
magnitude of the fluctuation also grows as the Universe evolves.

A nonzero variance of $\lvert H \rvert$ reflects the spatial inhomogeneity of the
scalar-doublet field. To understand why
$\lvert H(x^0,\mathbf{x}) \rvert$ may differ at distinct spatial points, we
revisit the expression of $H(x)$ in Eq.~\eqref{eq:m:HBR}, where the scalar-doublet field is
written as a stochastic integral over the causal support of the spacetime point
$x$. Consider two spacetime points with identical temporal coordinates but
different spatial locations, $x_1 = (x^0,\mathbf{x}_1)$, $x_2 = (x^0,\mathbf{x}_2)$,
and define their spatial separation as
$r = \lvert \mathbf{x}_1 - \mathbf{x}_2 \rvert$.
Figure~\ref{fig:Hs} illustrates the corresponding supporting regions
$D^{-}(x_1)$ and $D^{-}(x_2)$. These two regions are generally distinct. Since the
scalar-doublet field is constructed as a sum of independent random variables $dW(y)$
over the supporting domain, it follows that $H(x_1)$ and $H(x_2)$ are, in general,
different random variables.

When the distance $r$ is small compared with the conformal time $x^0$, the
supporting regions $D^{-}(x_1)$ and $D^{-}(x_2)$ have a large overlap (the heavily
shaded region in Fig.~\ref{fig:Hs}), while the non-overlapping parts,
$D^{-}(x_1)\cap \overline{D^{-}(x_2)}$ and
$\overline{D^{-}(x_1)}\cap D^{-}(x_2)$, are relatively small. In this regime, the
difference between $H(x_1)$ and $H(x_2)$ is expected to be minor. As the separation
$r$ becomes comparable to $x^0$, the overlap between $D^{-}(x_1)$ and $D^{-}(x_2)$
shrinks significantly relative to their total volumes, and the difference
between $H(x_1)$ and $H(x_2)$ correspondingly becomes more pronounced. For
$r > 2x^0$, the two supporting regions no longer overlap at all, implying that
$H(x_1)$ and $H(x_2)$ become statistically independent. Therefore, the
correlation length associated with the fluctuating scalar-doublet field at conformal time $x^0$ 
is expected to be of order $cx^0$.

Although the above argument is formulated for the individual components
of the scalar-doublet field, it extends naturally to the magnitude
$\lvert H \rvert$, which is a function of these components. In the
present work, we restrict ourselves to a qualitative analysis of the
spatial fluctuations of $\lvert H \rvert$ and do not attempt to derive
detailed observational consequences.

It is important to clarify the interpretation of the estimated relative
fluctuation,
$\Delta \lvert H \rvert / \langle \lvert H \rvert \rangle \sim 0.36$.
Within the present model, this quantity characterizes fluctuations of
the stochastic scalar-doublet field itself and should not be interpreted
as predicting observable variations on laboratory, planetary, or
galactic scales. According to the above estimate, the corresponding
correlation length at the current epoch is of the order of the present
cosmological horizon size,
$\xi_H \sim c t_c$, where $t_c$ denotes the current conformal time.
Consequently, observations performed within regions much smaller than
$\xi_H$ effectively probe a single realization of the stochastic
background and are therefore expected to exhibit only negligible local
variations.

If the stochastic scalar doublet considered here were embedded into a
more complete electroweak framework, its horizon-scale fluctuations
might, in principle, induce corresponding fluctuations in effective
physical parameters through Yukawa couplings. Whether such effects could
lead to observable cosmological or astrophysical signatures depends on
a detailed phenomenological analysis. These questions lie beyond the scope of the
present work and are left for future investigation.

\section{Fermion Dynamics Induced by the Stochastic Scalar-Doublet Field}
\label{sec:fdH}

In this section, we investigate how the stochastic scalar-doublet field
constructed in the previous sections influences fermionic dynamics through
a Yukawa interaction. Our purpose is not to construct the full
electroweak Yukawa sector of the Standard Model, but rather to explore,
within the present toy model, how the two components of the stochastic
scalar-doublet solution affect the evolution of fermionic fields. Throughout
our analysis, the fermionic fields are treated as quantized operators,
whereas the scalar-doublet field is regarded as a classical stochastic
background.

This semiclassical treatment neglects quantum fluctuations of the
scalar-doublet field, or equivalently, quantum processes involving the
creation and annihilation of scalar excitations. Under this approximation,
the stochastic scalar-doublet field obtained from the Euler-Lagrange
equation can be regarded as an external background and inserted directly
into the fermionic action. As will be shown below, the tail component
$H^{(\theta)}$ contributes as an effective mass-like term in the
Yukawa model, while the light-cone component $H^{(\delta)}$ naturally
induces a colored noise acting on fermions.

To illustrate these effects, we introduce the same Yukawa interaction as
used in the Standard Model for the electron and electron neutrino. This
choice is adopted solely as a convenient mathematical example for coupling
a scalar doublet to fermions and should not be interpreted as a complete
description of the electroweak Yukawa sector. The corresponding action is
\begin{equation}
\label{eq:n:sy}
S_Y
=
- y \int d^4 x \,
\left[
\left( \bar{\nu}_L, \bar{e}_L \right)
\begin{pmatrix}
H_A \\ H_B
\end{pmatrix}
e_R
+ \text{h.c.}
\right],
\end{equation}
where $e_L$ and $e_R$ denote the left- and right-handed electron fields,
respectively, $\bar e$ is the Dirac adjoint of $e$,
$\nu_L$ is the left-handed neutrino field,
and $y$ is the Yukawa coupling constant.
The scalar-doublet field
$\left(H_A,H_B\right)^T$
is given by the solution of the Euler-Lagrange equation derived in
Sec.~\ref{sec:EL}, as expressed in
Eqs.~\eqref{eq:EL:HdHt} and~\eqref{eq:EL:2He}.

In Eq.~\eqref{eq:n:sy}, we have assumed a flat Minkowski spacetime with
$g_{\mu\nu} \simeq \eta_{\mu\nu}$.
This approximation is justified because we are interested in the evolution
of fermions near the current epoch, $x^0 \simeq t_c$,
where spacetime curvature is negligible.
Moreover, the coordinate choice adopted in Sec.~\ref{sec:EL} ensures that
the metric reduces to $\eta_{\mu\nu}$ at $x^0 \simeq t_c$.

Combining the Yukawa action~\eqref{eq:n:sy} with the fermionic kinetic term,
namely the Dirac action
\(
- \int d^4 x \, \bar{\psi} \gamma^\mu \partial_\mu \psi,
\)
we obtain a theory describing fermions propagating in an external scalar-field background.
The quantization of this theory is straightforward.
Although the salar-doublet field $\left(H_A, H_B\right)$ is now a random-valued function
of spacetime, the Yukawa interaction contains no derivatives of the fermionic fields
and therefore acts purely as a potential term, without affecting the canonical
quantization procedure.

Applying the Legendre transformation in the usual manner, we obtain the Hamiltonian
in the Schr\"{o}dinger picture,
\begin{equation}
\label{eq:n:eH}
\begin{split}
\hat{\mathcal{H}}(t)
=
\hat{\mathcal{H}}_F
+
y \int d^3 \mathbf{x} \,
\left(
\hat{\bar{\nu}}_L(\mathbf{x}),
\hat{\bar{e}}_L(\mathbf{x})
\right)
\begin{pmatrix}
H_A(t,\mathbf{x}) \\
H_B(t,\mathbf{x})
\end{pmatrix}
\hat{e}_R(\mathbf{x})
+ \text{h.c.},
\end{split}
\end{equation}
where $t \equiv x^0$ denotes the time coordinate.
Here, hatted symbols represent operators, such as fermionic field operators.
The operator $\hat{\mathcal{H}}_F$ is the free Dirac Hamiltonian without a mass term
and is time independent.
All time dependence of the full Hamiltonian $\hat{\mathcal{H}}(t)$ arises from the scalar-doublet field,
which acts as a random, time-dependent external potential.

Let us consider the fermionic evolution governed by the Hamiltonian $\hat{\mathcal{H}}(t)$
over a finite time interval of duration $T$.
Without loss of generality, we take the evolution to occur from
$t_c$ to $t_c + T$, with $T \ll t_c$.
To isolate the effect of the scalar-doublet field on fermion dynamics,
we focus on the interaction term in Eq.~\eqref{eq:n:eH}.
The corresponding evolution operator is given by
\begin{equation}
\label{eq:n:Uue}
\begin{split}
\hat{U}
&= \mathcal{T}
\exp\left\{
- i \int_{t_c}^{t_c + T} dt \, \hat{\mathcal{H}}(t)
\right\}
\\
&\simeq
\exp \Bigg\{
- i y \int d^3 \mathbf{x} \,
\left(
\hat{\bar{\nu}}_L(\mathbf{x}),
\hat{\bar{e}}_L(\mathbf{x})
\right)
\begin{pmatrix}
\Theta_A(T,\mathbf{x}) \\
\Theta_B(T,\mathbf{x})
\end{pmatrix}
\hat{e}_R(\mathbf{x})
+ \text{h.c.}
\Bigg\},
\end{split}
\end{equation}
where $\mathcal{T}$ denotes the time-ordering operator.
In deriving the second line, we have neglected the free Dirac Hamiltonian
$\hat{\mathcal{H}}_F$, which only generates the standard free evolution of
fermions and does not affect the interaction effects arising from the
stochastic scalar-doublet field discussed here. The quantities
\begin{equation}
\Theta_{A/B}(T,\mathbf{x})
=
\int_{t_c}^{t_c + T} dt \,
H_{A/B}(t,\mathbf{x})
\end{equation}
represent the phases accumulated by fermions during the evolution as a result
of their interaction with the scalar-doublet field.
It is therefore evident that, over the time interval $[t_c, t_c + T]$,
the influence of the scalar-doublet field on fermionic evolution is completely determined
by the statistical properties of $\Theta(T,\mathbf{x})$.

\begin{figure}[tbp]
\vspace{0.2cm}
%\vspace{1mm}.
\includegraphics[width=0.8\linewidth]{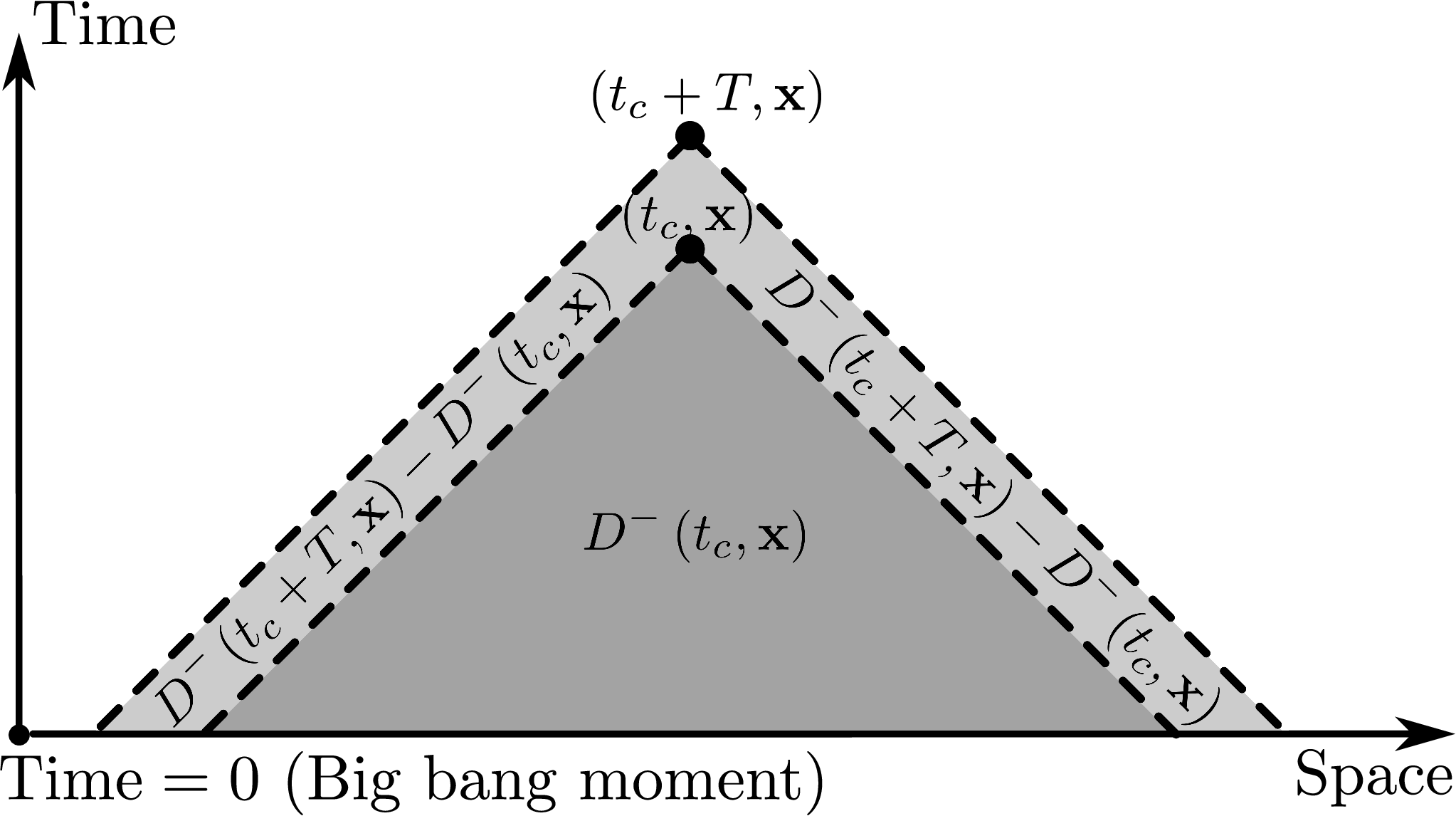}
\caption{Supporting regions of the scalar-doublet field at spacetime points
\(x=(x^0,\mathbf{x})\), where \(x^0\) spans the evolution interval
\((t_c,\,t_c+T)\).
The two black dots indicate the endpoints at \(x^0=t_c\) and \(x^0=t_c+T\),
respectively.
The heavily shaded region represents the common intersection of all supporting
regions \(D^-(x)\) for \(x^0\in(t_c,\,t_c+T)\),
while the lightly shaded region denotes the difference
\(F^-(\mathbf{x}) = D^-(t_c+T,\mathbf{x}) - D^-(t_c,\mathbf{x})\).}\label{fig:Hno}
\end{figure}

According to Eqs.~\eqref{eq:EL:HdHt} and~\eqref{eq:EL:2He}, the accumulated phase
$\Theta(T,\mathbf{x})$ can be expressed as a stochastic integral,
\begin{equation}
\label{eq:n:T2}
\begin{split}
& \Theta(T,\mathbf{x})
= \frac{\gamma}{4\pi}
\int d\Omega(y)\,
\left(-g(y)\right)^{\frac{1}{4}} \\ &
\int_{t_c}^{t_c+T} dx^0 \,
\Big[
U(x,y)\,\delta_+\!\left(\sigma(x,y)\right)
+ V(x,y)\,\theta_+\!\left(\sigma(x,y)\right)
\Big].
\end{split}
\end{equation}
The stochastic integration with respect to $d\Omega(y)$ extends over the
supporting region of $x=(x^0,\mathbf{x})$, namely
$C^-(x)\cup D^-(x)$, with $x^0\in(t_c,t_c+T)$.
As illustrated in Fig.~\ref{fig:Hno}, the union of these supporting regions
for all $x^0\in(t_c,t_c+T)$ is entirely contained within
$D^-(t_c+T,\mathbf{x})$.

To proceed, we decompose the integration domain into two nonintersecting regions,
$D^-(t_c+T,\mathbf{x}) = D^-(t_c,\mathbf{x}) \;\cup\; F^-(\mathbf{x})$, $F^-(\mathbf{x}) \equiv
D^-(t_c+T,\mathbf{x}) - D^-(t_c,\mathbf{x})$,
as shown schematically in Fig.~\ref{fig:Hno}.
This decomposition is crucial for separating the effects of the
$\delta_+$-term and the $\theta_+$-term in Eq.~\eqref{eq:n:T2}.
Correspondingly, the phase $\Theta(T,\mathbf{x})$ can be written as
\begin{equation}
\Theta(T,\mathbf{x})
=
\Theta^{(m)}(T,\mathbf{x})
+
\Theta^{(n)}(T,\mathbf{x}),
\end{equation}
where
\begin{equation}
\label{eq:n:Tm}
\Theta^{(m)}(T,\mathbf{x})
=
\frac{\gamma}{4\pi}
\int_{y\in D^-(t_c,\mathbf{x})}
d\Omega(y)\,
\left(-g(y)\right)^{\frac{1}{4}}
\int_{t_c}^{t_c+T} dx^0 \,
V(x,y)
\end{equation}
is identified as the mass term, since it contributes as an effective
mass-like contribution, and
\begin{equation}
\label{eq:n:Tn}
\begin{split}
& \Theta^{(n)}(T,\mathbf{x})
=
\frac{\gamma}{4\pi}
\int_{y\in F^-(\mathbf{x})}
d\Omega(y)\,
\left(-g(y)\right)^{\frac{1}{4}} \\ &
\int_{t_c}^{t_c+T} dx^0 \,
\Big[
U(x,y)\,\delta_+\!\left(\sigma(x,y)\right)
+ V(x,y)\,\theta_+\!\left(\sigma(x,y)\right)
\Big]
\end{split}
\end{equation}
is the noise term, since it generates a colored stochastic
contribution to the fermion phase.

In Eq.~\eqref{eq:n:Tm}, the $\delta_+$-term gives no contribution.
This is because the support of $\delta_+\!\left(\sigma(x,y)\right)$,
namely the set of points satisfying
$\lvert y^0-x^0\rvert = \lvert \mathbf{y}-\mathbf{x}\rvert$,
lies entirely outside $D^-(t_c,\mathbf{x})$ for all
$x^0\in(t_c,t_c+T)$.
By contrast, the entire region $D^-(t_c,\mathbf{x})$ is always contained
within the support of $\theta_+\!\left(\sigma(x,y)\right)$,
corresponding to
$\lvert y^0-x^0\rvert > \lvert \mathbf{y}-\mathbf{x}\rvert$.
Therefore, within $\Theta^{(m)}$ one may set
$\theta_+(\sigma)\equiv 1$.

It is important to emphasize that the mass term $\Theta^{(m)}$ and the noise term
$\Theta^{(n)}$ are stochastic integrals over nonoverlapping spacetime regions.
Since the noise field $d\Omega(y)$ is composed of independent random variables at
different spacetime points, it follows immediately that
$\Theta^{(m)}(T,\mathbf{x})$ and $\Theta^{(n)}(T,\mathbf{x})$
are statistically independent random quantities.
This decomposition is deliberately constructed to ensure such independence,
allowing the physical effects of the mass term and the noise term
to be analyzed separately.

\subsection{Effect of the Mass Term $\Theta^{(m)}$}
\label{sec:n:Tm}

Let us first study the effect of the mass term $\Theta^{(m)}$, defined in
Eq.~\eqref{eq:n:Tm}. We focus on evolution over time intervals
$T$ that are much shorter than the present conformal time $t_c$, which is
cosmologically large. When $T \ll t_c$, it is reasonable to approximate $V(x,y)$ in
Eq.~\eqref{eq:n:Tm}---for $x^0\in(t_c,t_c+T)$---by its value at $x^0=t_c$.
With this approximation, $\int_{t_c}^{t_c+T} dx^0 \, V(x,y)
\;\approx\;
T\,V(x,y)\big|_{x^0=t_c}$.
By comparing the resulting expression for $\Theta^{(m)}(T,\mathbf{x})$
with the definition of $H^{(\theta)}(t_c,\mathbf{x})$ in
Eq.~\eqref{eq:EL:2He}, we obtain
\begin{equation}
\label{eq:n:Tma}
\Theta^{(m)}(T,\mathbf{x})
\;\approx\;
T\,H^{(\theta)}(t_c,\mathbf{x}).
\end{equation}

The approximation~\eqref{eq:n:Tma} relies on the fact that
$H^{(\theta)}(x^0,\mathbf{x})$ varies slowly with respect to $x^0$.
As shown in Sec.~\ref{sec:EH}, the standard deviation of
$H^{(\theta)}(x^0,\mathbf{x})$ grows linearly with $x^0$.
Therefore, for cosmologically large $x^0$ and for $T\ll t_c$, the variation
of $H^{(\theta)}$ over the interval $(t_c,t_c+T)$ is negligible, and
$H^{(\theta)}(x^0,\mathbf{x})$ may be replaced by its value at $x^0=t_c$.
Since $\Theta^{(m)}$ is precisely the time integral of
$H^{(\theta)}(x^0,\mathbf{x})$ over this interval,
Eq.~\eqref{eq:n:Tma} provides an accurate approximation.

Equation~\eqref{eq:n:Tma} shows that the phase $\Theta^{(m)}$ is uniquely
determined by $H^{(\theta)}$ and is independent of $H^{(\delta)}$.
The statistical properties of $H^{(\theta)}$ were studied in
Sec.~\ref{sec:Htt}. In particular, there always exists a local
$\mathrm{U(1)}\times\mathrm{SU(2)}$ transformation $\mathcal{U}$ that
eliminates $H^{(\theta)}_A$ together with the imaginary part of
$H^{(\theta)}_B$, leaving only the real part of $H^{(\theta)}_B$
(see Eq.~\eqref{eq:m:HctH}).
Since $\Theta^{(m)}$ is proportional to $H^{(\theta)}$, under this choice of
gauge it becomes
\begin{equation}
\label{eq:n:Tme}
\Theta^{(m)}
\stackrel{\mathcal{U}}{\longrightarrow}
\begin{pmatrix}
0 \\ T\lvert H\rvert
\end{pmatrix}.
\end{equation}

Substituting Eq.~\eqref{eq:n:Tme} into Eq.~\eqref{eq:n:Uue}, the evolution
operator generated by $\Theta^{(m)}$ takes the form
\begin{equation}
\label{eq:n:Ume}
\hat{U}^{(m)}
=
\exp\!\left\{
-i T y \int d^3\mathbf{x}\,
\lvert H(t_c,\mathbf{x})\rvert
\left(
\hat{\bar e}_L(\mathbf{x})\hat e_R(\mathbf{x})
+ h.c.
\right)
\right\}.
\end{equation}
As discussed in Sec.~\ref{sec:HFF}, the spatial fluctuation of
$\lvert H(t_c,\mathbf{x})\rvert$ is negligible between two points separated
by a distance $\lvert\mathbf{x}_1-\mathbf{x}_2\rvert\ll t_c$.
For any local fermionic process, the spatial extent of the fermionic wave
packet is negligibly small compared with the cosmological scale $t_c$.
We may therefore treat $\lvert H(t_c,\mathbf{x})\rvert$ as spatially uniform,
or further replace it by its expectation value
$\langle\lvert H(t_c)\rvert\rangle$, defined in
Eq.~\eqref{eq:m:Hva}.
The evolution operator then simplifies to
\begin{equation}
\label{eq:n:Umf}
\hat{U}^{(m)}
=
\exp\!\left\{
-i T y \langle\lvert H(t_c)\rvert\rangle
\int d^3\mathbf{x}\,
\hat{\bar\psi}_e(\mathbf{x})\hat\psi_e(\mathbf{x})
\right\},
\end{equation}
where $\psi_e = e_L + e_R$ denotes the electron field.

Equation~\eqref{eq:n:Umf} has precisely the form of the evolution operator for
a massive Dirac fermion. Within the present toy model, it is therefore natural
to identify the effective fermion mass as
\begin{equation}
\label{eq:n:me}
m_e
=
y\,\langle\lvert H(t_c)\rvert\rangle
=
\frac{3\gamma\eta y\,t_c}{8\sqrt{2}}.
\end{equation}
This result explains why $\Theta^{(m)}$ is referred to as the mass term: its
effect on fermionic evolution is mathematically equivalent to that generated by
a fermion mass. In the present framework, the effective mass arises from the
ensemble-average magnitude of the stochastic scalar-doublet field rather than
from the vacuum expectation value of the Standard Model Higgs field.
Equation~\eqref{eq:n:me} establishes the correspondence between the parameters
of the toy model and the effective fermion mass that emerges from its
stochastic dynamics.

\subsection{Properties of the Noise Term}
\label{sec:pnt}

Next, we study the statistical properties of the noise term
$\Theta^{(n)}$, defined in Eq.~\eqref{eq:n:Tn}.
In the previous analysis, we emphasized that one must choose a gauge---or
equivalently a rotation $\mathcal{U}$ acting on the scalar doublet---to
eliminate the first component of $H^{(\theta)}$ (and hence
$\Theta^{(m)}$) and render the second component real.
Such a rotation is uniquely determined by $\Theta^{(m)}$.
However, since $\mathcal{U}$ acts on the entire scalar doublet, it also
acts simultaneously on $\Theta^{(n)}$.

Fortunately, $\Theta^{(m)}$ and $\Theta^{(n)}$ are defined as stochastic
integrals over nonintersecting spacetime domains, and are therefore
statistically independent random variables.
As a result, the rotation $\mathcal{U}$ determined by $\Theta^{(m)}$ is
independent of $\Theta^{(n)}$.
Moreover, $\Theta^{(n)}$, like the scalar-doublet field itself, is statistically
invariant under $\mathrm{U(1)}\times\mathrm{SU(2)}$ transformations.
Consequently, the action of $\mathcal{U}$ has no effect on the
statistical properties of $\Theta^{(n)}$, and can be ignored
in the following analysis.

According to Eq.~\eqref{eq:n:Tn}, $\Theta^{(n)}$ consists of two
contributions: one from the $U\delta_+$ term and the other from the
$V\theta_+$ term, originating from $H^{(\delta)}$ and $H^{(\theta)}$,
respectively.
We denote these contributions by $\Theta^{(n,\delta)}$ and
$\Theta^{(n,\theta)}$.
We will study them separately.
As will be shown below, the dominant contribution comes from
$\Theta^{(n,\delta)}$, while the contribution from $\Theta^{(n,\theta)}$
is subleading and can be neglected.

Explicitly, $\Theta^{(n,\delta)}$ can be written as
\begin{equation}
\label{eq:n:Tnd}
\Theta^{(n,\delta)}
=
\frac{\gamma}{4\pi}
\int_{t_c}^{t_c+T} dx^0
\int_{y\in F^-} d\Omega(y)\,
\left(-g(y)\right)^{\frac{1}{4}}
U(x,y)\,\delta_+\!\left(\sigma(x,y)\right).
\end{equation}
As illustrated in Fig.~\ref{fig:Hno}, the integration domain $F^-$
corresponds precisely to the set of past-directed null geodesics from
$x=(x^0,\mathbf{x})$ with $x^0\in(t_c,t_c+T)$.
In other words, $F^-$ coincides with the supporting domain of
$\delta_+\!\left(\sigma(x,y)\right)$, namely the set of points $y$
satisfying $\sigma(x,y)=0$.
Therefore, explicitly restricting the integration domain to $F^-$ is
redundant.
Replacing $\int_{y\in F^-} d\Omega(y)$ by $\int d\Omega(y)$ in
Eq.~\eqref{eq:n:Tnd} does not change the result.

After this replacement, it becomes immediately clear that
$\Theta^{(n,\delta)}$ has the same mathematical structure as the phase
accumulated by a fermion evolving under the colored noise
$h(x)$ introduced in our previous work~\cite{Wang24}.
In particular, if one chooses $U(x,y)\equiv 1$, corresponding to the
flat-spacetime solution (see Eq.~\eqref{eq:EL:Ua}), then
$\Theta^{(n,\delta)}$ is exactly equal to the phase generated by $h(x)$,
whose statistical properties have already been studied in detail.

This correspondence is not accidental.
Indeed, if we assume flat spacetime ($g_{\mu\nu}=\eta_{\mu\nu}$) and set
$\eta=0$---which is equivalent to neglecting $H^{(\theta)}$ or,
equivalently, the $V(x,y)$ term proportional to $\eta^2$---then the
Euler-Lagrange equation~\eqref{eq:EL} reduces precisely to the
D'Alembert equation satisfied by $h(x)$.
It is therefore natural that $H^{(\delta)}$ and the associated noise
term $\Theta^{(n,\delta)}$ inherit the same statistical properties as
$h(x)$.

For the sake of self-consistency, we briefly repeat here the analysis
presented in Ref.~[\onlinecite{Wang24}], in order to derive the statistical properties of
$\Theta^{(n,\delta)}$. First, $\Theta^{(n,\delta)}$ is defined as a stochastic integral over
$d\Omega(y)$, whose values at different spacetime points $y$ are
independent Gaussian random variables.
Consequently, $\Theta^{(n,\delta)}$, being a linear combination of such
variables, must itself obey a Gaussian distribution.
Moreover, since $d\Omega(y)$ has vanishing expectation value, it follows
immediately that $\left\langle \Theta^{(n,\delta)} \right\rangle = 0 $.
The remaining task is to determine the variance and, more generally, the
spatial correlation properties of $\Theta^{(n,\delta)}$.
To this end, we introduce the spatial correlation function
\begin{equation}
\label{eq:n:Co}
C(r)
=
\left\langle
\Theta^{(n,\delta)}_{BR}\!\left(T,\mathbf{x}_1\right)\,
\Theta^{(n,\delta)}_{BR}\!\left(T,\mathbf{x}_2\right)
\right\rangle ,
\end{equation}
where $r=\left|\mathbf{x}_1-\mathbf{x}_2\right|$ denotes the spatial
separation between the two points.

Here we have chosen the specific component $\Theta_{BR}$ for definiteness.
However, due to the $\mathrm{U(1)}\times\mathrm{SU(2)}$ invariance of the
noise, the correlation functions of all four real components of
$\Theta^{(n,\delta)}$ are identical.
According to the definition~\eqref{eq:n:Co}, the variance of
$\Theta^{(n,\delta)}$ is simply given by $C(r)$ evaluated at $r=0$.

At the same time, the behavior of $C(r)$ for $r>0$ characterizes the
spatial fluctuation of $\Theta^{(n,\delta)}(T,\mathbf{x})$.
If the spatial fluctuation is weak, the correlation function $C(r)$
decays slowly as $r$ increases.
Conversely, a rapid decay of $C(r)$ with $r$ indicates strong spatial
fluctuations of the noise.

\begin{figure}[tbp]
\vspace{0.2cm}
%\vspace{1mm}.
\includegraphics[width=0.8\linewidth]{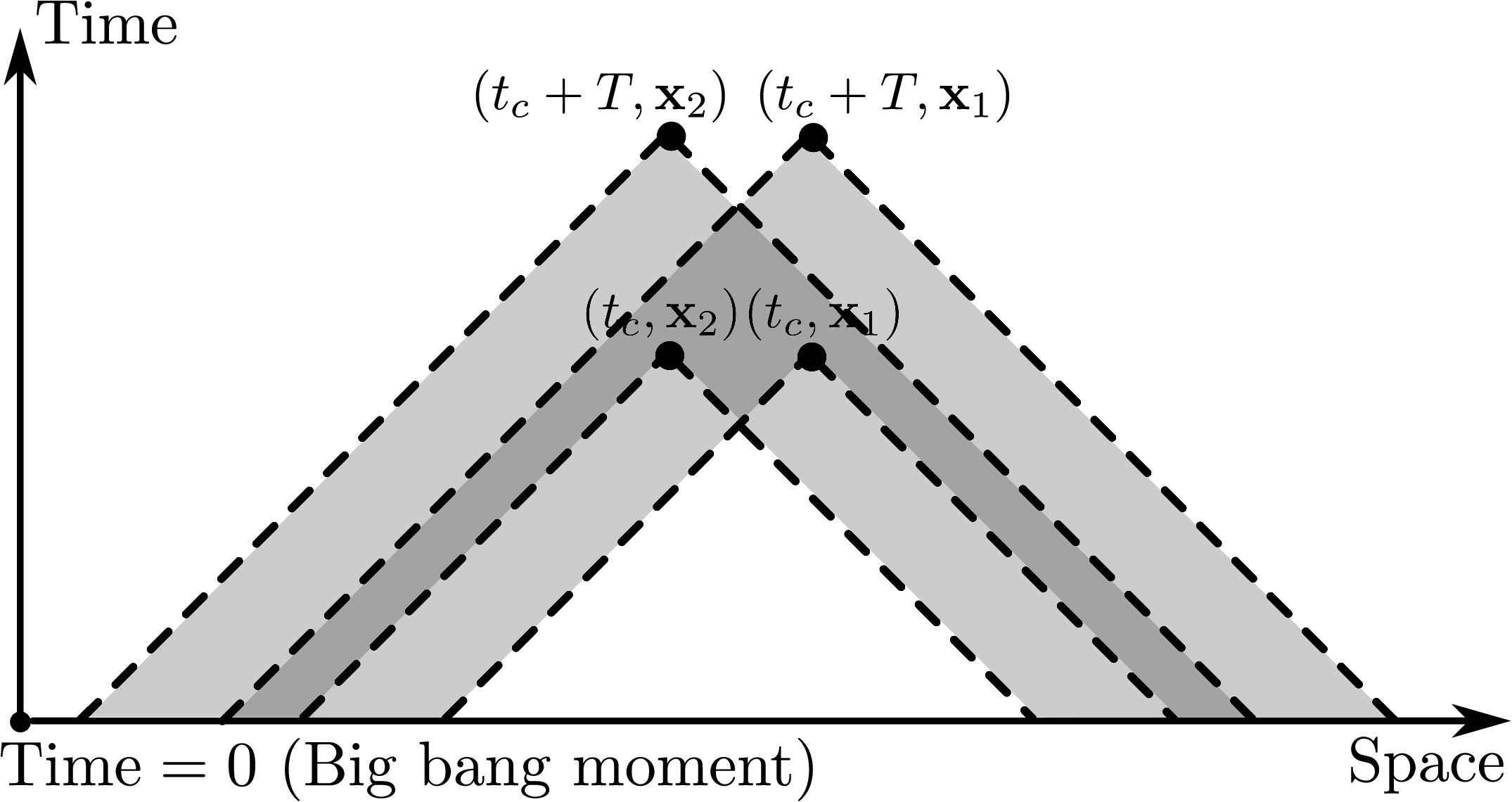}
\caption{Schematic diagram of the supporting regions \(F^-(\mathbf{x}_1)\) and
\(F^-(\mathbf{x}_2)\) at two distinct spatial locations, indicated by the shaded areas.
The heavily shaded region represents the intersection of the two supporting regions.}\label{fig:HCc}
\end{figure}

We now calculate the correlation function $C(r)$, following the approach of
Ref.~[\onlinecite{Wang24}].
For convenience, we adopt the flat-spacetime approximation discussed above,
and set the scale factor $a\equiv 1$.
This choice considerably simplifies the expressions of $g(y)$, $U(x,y)$,
and $\sigma(x,y)$. Using the identity $\delta_+\left(\sigma(x,y)\right) =
\left. {\delta\left(x^0-y^0-\left|\mathbf{x}-\mathbf{y}\right|\right)}
\middle/ {\left|\mathbf{x}-\mathbf{y}\right|} \right.$,
and the statistical properties of $d\Omega(y)$, we obtain
\begin{equation}
\label{eq:n:CrC}
C(r)
=
\frac{\gamma^2}{(4\pi)^2}
\int_{y \in F^-(\mathbf{x}_1)\cap F^-(\mathbf{x}_2)}
d^4 y\,
\frac{1}
{\left|\mathbf{x}_1-\mathbf{y}\right|
 \left|\mathbf{x}_2-\mathbf{y}\right|},
\end{equation}
where the integration domain is the intersection between
$F^-(\mathbf{x}_1)$ and $F^-(\mathbf{x}_2)$, namely the respective
supporting regions of
$\Theta^{(n,\delta)}(T,\mathbf{x}_1)$ and
$\Theta^{(n,\delta)}(T,\mathbf{x}_2)$.
Figure~\ref{fig:HCc} schematically illustrates this intersection region
(the heavily shadowed area).

From the geometry of the integration domain, it is clear that its shape
depends on whether $r<T$ or $r>T$.
Therefore, in principle, $C(r)$ must be evaluated separately in these two
regimes. In the present work, however, we restrict attention to the regime
$r\ll T$. Indeed, adopting the natural unit $c=1$, a typical laboratory timescale
(e.g.\ $T\sim10^{-8}\,\mathrm{s}$) corresponds to a macroscopic distance
($\sim3\,\mathrm{m}$), while the spatial extension of a fermionic wave packet
is usually much smaller.

Although the integral in Eq.~\eqref{eq:n:CrC} cannot be evaluated
analytically for general $r$, in the regime $r\ll T$ we may expand the
integrand as a power series in $r$ and keep only the leading terms.
Carrying out this expansion and retaining terms up to first order in $r$,
we obtain
\begin{equation}
\label{eq:n:Crre}
C(r)
\approx
\frac{\gamma^2}{4\pi}\, t_c
\left(
T - \frac{r}{2}
\right).
\end{equation}
The variance of $\Theta^{(n,\delta)}$ is therefore given by
\begin{equation}\label{eq:n:Vtex}
\mathrm{Var}\!\left(\Theta^{(n,\delta)}\right)
=
C(0)
=
\frac{\gamma^2}{4\pi}\, t_c T .
\end{equation}
Equation~\eqref{eq:n:Crre} further shows that the spatial correlation decays
linearly with $r$, with a characteristic decay rate of
$(8\pi)/(\gamma^2 t_c)$.
Importantly, $C(r)$ is not proportional to a $\delta$-function of $r$,
demonstrating that the noise induced by $H^{(\delta)}$ is a colored noise
rather than a white one.

We now show that the contribution from $\Theta^{(n,\theta)}$ is much smaller
than that from $\Theta^{(n,\delta)}$ and can be safely neglected.
The phase $\Theta^{(n,\theta)}$ is given by
\begin{equation}
\label{eq:n:Tnth}
\Theta^{(n,\theta)}
=
\frac{\gamma}{4\pi}
\int_{t_c}^{t_c+T} dx^0
\int_{y\in F^-} d\Omega(y)\,
\left(-g(y)\right)^{1/4}
V(x,y)\,
\theta_+\!\left(\sigma\right).
\end{equation}
Like $\Theta^{(n,\delta)}$, $\Theta^{(n,\theta)}$ is a Gaussian random
variable with vanishing mean.
Thus, its physical relevance is entirely determined by its variance.
Proceeding analogously to the previous calculation, we obtain
\begin{equation}
\begin{split}
\mathrm{Var}\!\left(\Theta^{(n,\theta)}_{BR}\right)
=&\,
\frac{\gamma^2\eta^4}{(4\pi)^2}
\int_{y\in F^-(\mathbf{x})} d^4 y \\
&\times
\left(
\int_{y^0+\left|\mathbf{x}-\mathbf{y}\right|}^{t_c+T}
dx^0\,
\frac{J_1\!\left(\sqrt{2\eta^2\sigma(x,y)}\right)}
{\sqrt{2\eta^2\sigma(x,y)}}
\right)^2 .
\end{split}
\end{equation}
After an appropriate change of integration variables and some algebra, this
expression can be rewritten as
\begin{equation}
\begin{split}
\mathrm{Var}\!\left(\Theta^{(n,\theta)}_{BR}\right)
=&\,
\frac{\gamma^2}{2\pi\eta^2}
\int_0^{\eta t_c} d\rho
\int_0^{\sqrt{(\eta T)^2+2\eta T\rho}} ds_a
\int_0^{s_a} ds_b \\
&\times
J_1(s_a)J_1(s_b)
\frac{\rho^2
\left(\eta T+\rho-\sqrt{s_a^2+\rho^2}\right)}
{\sqrt{s_a^2+\rho^2}\sqrt{s_b^2+\rho^2}} .
\end{split}
\end{equation}

In the limiting case $\eta T\gg1$ (note that $\eta T$ is dimensionless), this
integral evaluates to
$\gamma^2 t_c T /(4\pi)$, coinciding with the variance of
$\Theta^{(n,\delta)}_{BR}$.
However, the parameter regime considered in Sec.~\ref{sec:para}
corresponds to $\eta T\ll1$ for the time intervals of interest.
In this regime, one finds $\mathrm{Var}\!\left(\Theta^{(n,\theta)}_{BR}\right) \ll
\mathrm{Var}\!\left(\Theta^{(n,\delta)}_{BR}\right)$. Both $\Theta^{(n,\theta)}_{BR}$
and $\Theta^{(n,\delta)}_{BR}$ follow Gaussian distribution with zero mean.
Therefore, $\Theta^{(n,\theta)}$ provides a negligible contribution to the
noise term, and we may safely approximate
\begin{equation}
\Theta^{(n)} \simeq \Theta^{(n,\delta)} .
\end{equation}
Consequently, the correlation function and variance given in
Eqs.~\eqref{eq:n:Crre} and~\eqref{eq:n:Vtex}, respectively, indeed characterizes the properties of the
entire noise term $\Theta^{(n)}$.

\subsection{Effect of the Noise Term}
\label{sec:effn}

We are now prepared to study the effect of the noise term on the evolution of
fermions. Using Eq.~\eqref{eq:n:Uue}, the evolution operator generated by
$\Theta^{(n)}$ can be written as
\be\label{eq:n:UTn}
\begin{split}
\hat{U}^{(n)} =
\exp \Bigg\{ &
-i \int d^3 \mathbf{x}
\left[
y\,\Theta_A^{(n)}(T,\mathbf{x})\,
\hat{\bar{\nu}}_L(\mathbf{x}) \hat{e}_R(\mathbf{x})
+ h.c.
\right] \\
&- i \int d^3 \mathbf{x}
\left[
y\,\Theta_B^{(n)}(T,\mathbf{x})\,
\hat{\bar{e}}_L(\mathbf{x}) \hat{e}_R(\mathbf{x})
+ h.c.
\right]
\Bigg\}.
\end{split}
\ee
Two independent noise-induced phases, $\Theta_A^{(n)}$ and $\Theta_B^{(n)}$,
are thus coupled to the fermionic bilinears $\hat{\bar{\nu}}_L \hat{e}_R$ and
$\hat{\bar{e}}_L \hat{e}_R$, respectively. Within the present toy model, the first term in
Eq.~\eqref{eq:n:UTn} couples the noise field to the bilinear
$\hat{\bar{\nu}}_L \hat{e}_R$ through the first component of the
scalar doublet. The consequences of this coupling are not investigated further in the
present work. Instead, we focus on the second term, which produces a colored stochastic
phase acting on the electron field and provides the main example analyzed below.

The phase $\Theta_B^{(n)}$ is complex-valued and can be decomposed as
$\Theta_B^{(n)}=\Theta_{BR}^{(n)}+i\Theta_{BI}^{(n)}$.
Correspondingly, the electron-noise coupling in
Eq.~\eqref{eq:n:UTn} can be rewritten as
\be\label{eq:n:UTn1}
\begin{split}
\hat{U}^{(n)} =
\exp \Bigg\{ &
-i \int d^3 \mathbf{x}\,
\bigl(y\,\Theta_{BR}^{(n)}(T,\mathbf{x})\bigr)\,
\hat{\bar{\psi}}_e(\mathbf{x}) \hat{\psi}_e(\mathbf{x}) \\
&+ \int d^3 \mathbf{x}\,
\bigl(y\,\Theta_{BI}^{(n)}(T,\mathbf{x})\bigr)
\left[
\hat{\bar{e}}_L(\mathbf{x}) \hat{e}_R(\mathbf{x})
- h.c.
\right]
\Bigg\},
\end{split}
\ee
where we have used the identity
$\hat{\bar{e}}_L \hat{e}_R + (\hat{\bar{e}}_L \hat{e}_R)^\dagger = \hat{\bar{\psi}}_e \hat{\psi}_e$,
with $\hat{\psi}_e=\hat{e}_L+\hat{e}_R$ being the Dirac electron field.

Both $\Theta_{BR}^{(n)}$ and $\Theta_{BI}^{(n)}$ contribute to the electron
dynamics.
However, their physical effects are not equally important.
To see this explicitly, we expand the fermionic fields in momentum space and
work in the Weyl basis.
The left- and right-handed components of the electron field read
\be\label{eq:n:eLR}
\begin{split}
\hat{e}_L(\mathbf{x}) =
\frac{1}{\sqrt{(2\pi)^3}}
&\int d^3\mathbf{p}\,
\frac{m_e+E_{\mathbf{p}}-\mathbf{p}\cdot\vec{\sigma}}
{\sqrt{2E_{\mathbf{p}}(m_e+E_{\mathbf{p}})}} \\
&\times
\left[
\frac{e^{i\mathbf{p}\cdot\mathbf{x}}}{\sqrt{2}}
\begin{pmatrix}
\hat{c}_{\mathbf{p}\uparrow}\\
\hat{c}_{\mathbf{p}\downarrow}\\
0\\
0
\end{pmatrix}
+
\frac{e^{-i\mathbf{p}\cdot\mathbf{x}}}{\sqrt{2}}
\begin{pmatrix}
-\hat{d}^\dagger_{\mathbf{p}\downarrow}\\
\hat{d}^\dagger_{\mathbf{p}\uparrow}\\
0\\
0
\end{pmatrix}
\right], \\[1ex]
\hat{e}_R(\mathbf{x}) =
\frac{1}{\sqrt{(2\pi)^3}}
&\int d^3\mathbf{p}\,
\frac{m_e+E_{\mathbf{p}}+\mathbf{p}\cdot\vec{\sigma}}
{\sqrt{2E_{\mathbf{p}}(m_e+E_{\mathbf{p}})}} \\
&\times
\left[
\frac{e^{i\mathbf{p}\cdot\mathbf{x}}}{\sqrt{2}}
\begin{pmatrix}
0\\
0\\
\hat{c}_{\mathbf{p}\uparrow}\\
\hat{c}_{\mathbf{p}\downarrow}
\end{pmatrix}
+
\frac{e^{-i\mathbf{p}\cdot\mathbf{x}}}{\sqrt{2}}
\begin{pmatrix}
0\\
0\\
\hat{d}^\dagger_{\mathbf{p}\downarrow}\\
-\hat{d}^\dagger_{\mathbf{p}\uparrow}
\end{pmatrix}
\right],
\end{split}
\ee
where $E_{\mathbf{p}}=\sqrt{m_e^2+\mathbf{p}^2}$,
$\vec{\sigma}$ denotes the Pauli matrices, and
$\hat{c}_{\mathbf{p}s}$ and $\hat{d}_{\mathbf{p}s}$
($s=\uparrow,\downarrow$) are the annihilation operators of electrons and
positrons, respectively.

We have explicitly included the electron mass $m_e$ in
Eq.~\eqref{eq:n:eLR}. As shown in Sec.~\ref{sec:n:Tm}, the tail component
$H^{(\theta)}$ contributes to the fermion dynamics through an effective
mass term, with $m_e$ defined by Eq.~\eqref{eq:n:me}.
Therefore, when analyzing the effect of $H^{(\delta)}$, the contribution
of this effective mass term must be included consistently.

Using Eq.~\eqref{eq:n:eLR}, one finds that the scalar bilinear
$\hat{\bar{e}}_L \hat{e}_R + (\hat{\bar{e}}_L \hat{e}_R)^\dagger = \hat{\bar{\psi}}_e \hat{\psi}_e$
is parametrically much larger than the combination
$\hat{\bar{e}}_L \hat{e}_R - (\hat{\bar{e}}_L \hat{e}_R)^\dagger$.
This becomes particularly transparent in the nonrelativistic limit,
where $E_{\mathbf{p}}\approx m_e$ and the momentum-dependent terms
$\mathbf{p}\cdot\vec{\sigma}$ can be neglected.
In this limit, one finds $\hat{\bar{e}}_L \hat{e}_R - (\hat{\bar{e}}_L \hat{e}_R)^\dagger = 0$,
while $\hat{\bar{\psi}}_e \hat{\psi}_e$ reduces to the nonrelativistic electron density
operator. Consequently, the contribution from $\Theta_{BI}^{(n)}$ is negligible in the
nonrelativistic regime.

Let us consider the evolution of a single-electron wave packet under the action
of $\hat{U}^{(n)}$.
For simplicity, we neglect the kinetic term in the Hamiltonian.
Assuming the initial wave function at time $t_c$ to be
$\Psi_i(\mathbf{x})$, then in the nonrelativistic limit the final wave function
at time $t_c+T$ is given by
\be\label{eq:n:PfPi}
\Psi_f(\mathbf{x})
=
\exp\!\left\{
-i\,y\,\Theta_{BR}^{(n)}(T,\mathbf{x})
\right\}
\Psi_i(\mathbf{x}),
\ee
where we have used the properties of the density operator
$\hat{\bar{\psi}}_e\hat{\psi}_e$. 
Strictly speaking, the discussion should be formulated entirely in terms of
second-quantized operators.
Here, the single-particle wave function is introduced as a convenient
representation of the one-electron sector of the Fock space.
All results can be equivalently derived by evaluating the action of
$\hat{U}^{(n)}$ on one-particle states.

Equation~\eqref{eq:n:PfPi} makes explicit why $H^{(\delta)}$ and the associated
phase $\Theta^{(n)}$ can be interpreted as a noise term.
The effect of $H^{(\delta)}$ is to generate an additional
$\mathbf{x}$-dependent phase on the wave function, with
$y\,\Theta_{BR}^{(n)}(T,\mathbf{x})$ being a Gaussian random variable of zero mean.
This is precisely the effect of an external noise potential coupled to the
density operator, which produces random phases in the wave function.

The statistical properties of this random phase have been derived above.
We now consider the phase difference between two spatial points
$\mathbf{x}_1$ and $\mathbf{x}_2$, separated by a distance
$r = |\mathbf{x}_1-\mathbf{x}_2|$.
The phase difference $\left( y\,\Theta_{BR}^{(n)}(T,\mathbf{x}_1) - y\,\Theta_{BR}^{(n)}(T,\mathbf{x}_2)\right)$
is itself a Gaussian random variable with zero mean.
Using Eq.~\eqref{eq:n:Crre}, its variance is found to be
\be
\begin{split}
\Big\langle
\big(
y\,\Theta_{BR}^{(n)}(T,\mathbf{x}_1)
-
y\,\Theta_{BR}^{(n)}(T,\mathbf{x}_2)
\big)^2
\Big\rangle
=
\frac{y^2\gamma^2 t_c}{4\pi}\, r .
\end{split}
\ee

For a Gaussian random variable with vanishing mean, the variance characterizes
the typical magnitude of its fluctuations.
We see that the variance of the phase difference grows linearly with the spatial
separation $r$ and becomes of order unity at the distance
\be\label{eq:n:xixi}
\xi = \frac{4\pi}{y^2\gamma^2 t_c}.
\ee
Since an arbitrary phase difference can always be mapped into the interval
$[0,2\pi)$, a fluctuation of order $\mathcal{O}(1)$ represents a significant
phase difference.
By contrast, when the variance is much smaller than unity, the phase difference
can be regarded as negligible.
Therefore, the length scale $\xi$ defined in Eq.~\eqref{eq:n:xixi} indeed
characterizes the distance beyond which the phase difference becomes
statistically significant, and can be identified as the \emph{phase correlation
length}. The appearance of the parameter $t_c$ in the correlation length $\xi$ reflects the
fact that $\Theta^{(n)}$ accumulates over the entire past light cone of the
evolution interval. 

In this sense, the noise $H^{(\delta)}$ induces random phases at different spatial
points, with correlations persisting up to the length scale $\xi$.
This behavior is characteristic of a colored noise.
By contrast, for a white noise the phase difference between two points would be
independent of their spatial separation.

We emphasize that the kinetic term in the Hamiltonian has been neglected here solely for the purpose
of isolating the effect of the noise-induced phase.
Including the kinetic Hamiltonian would lead to the usual spatial spreading and
dispersion of the wave packet, but would not alter the local phase modulation
generated by $\Theta^{(n)}$.

We have thus shown that the two components of the scalar-doublet field,
$H^{(\theta)}$ and $H^{(\delta)}$, influence fermion dynamics in
qualitatively different ways.
Within the present framework, $H^{(\theta)}$ contributes an effective
mass term, whereas $H^{(\delta)}$ induces a colored stochastic phase
acting on the fermionic degrees of freedom.
The statistical properties of this colored noise, including its
correlation function and correlation length, follow directly from the
stochastic dynamics of the scalar-doublet field.
The colored-noise coupling obtained here provides a field-theoretic
realization of the phenomenological coupling introduced in
Ref.~\cite{Wang24}.

\section{Illustrative Parameter Estimates and Consistency of the Approximations}
\label{sec:para}

In this section, we present illustrative estimates of the model parameters
$\eta$ and $\gamma$. Our purpose is not to determine these parameters
uniquely, but rather to demonstrate that the model admits a physically
reasonable parameter regime in which the approximations employed in the
preceding sections are mutually consistent. In particular, we show that
there exist parameter values satisfying both
$\eta T < 1$ and $\eta t_c \gg 1$, which are the conditions required for
the validity of the analytical approximations adopted throughout this work.

As an illustrative example, we combine
Eqs.~\eqref{eq:n:me} and~\eqref{eq:n:xixi}. Restoring SI units, these
equations become
\begin{equation}
\label{eq:p:ee}
\begin{split}
\left\{
\begin{array}{c}
m_e
=
\displaystyle
\frac{3(y\gamma)c t_c}{8\sqrt{2}}
\,\eta,
\\[6pt]
\xi
=
\displaystyle
\frac{4\pi}
{(y\gamma)^2 c t_c},
\end{array}
\right.
\end{split}
\end{equation}
where $\eta$ has the dimension of mass, $\gamma$ has the dimension of inverse length,
$y$ is dimensionless, and $c$ denotes the speed of light.
Here, $m_e$ is the electron mass, while $\xi$ is the phase correlation length.
Equation~\eqref{eq:p:ee} provides two relations among
$\eta$, $\gamma$, $m_e$, $\xi$, and $t_c$.
Given illustrative choices of the physical quantities
$m_e$, $\xi$, and $t_c$, these relations yield corresponding values of
the model parameters.

Solving Eq.~\eqref{eq:p:ee} for $\eta$, we obtain
\begin{equation}
\label{eq:p:eta}
\eta
=
\frac{4\sqrt{2}}{3\sqrt{\pi}}
\sqrt{\frac{\xi}{c t_c}}
\,m_e .
\end{equation}
At the current cosmic epoch, the conformal time can be roughly estimated as
$t_c\sim10^{10}$ years, and the electron mass is
$m_e\approx9.1\times10^{-31}\,\mathrm{kg}$.
The parameter $\xi$ characterizes the spatial correlation length of the
noise-induced phase derived in Sec.~\ref{sec:effn}.
If such a stochastic phase contributes to the gradual loss of quantum
coherence, its associated correlation length would naturally be expected
to lie in a mesoscopic regime, separating microscopic systems that
preserve coherence from macroscopic systems in which coherence is rapidly
suppressed. Motivated by this general expectation, which is common to
many stochastic collapse and decoherence models, we adopt the
illustrative value
$\xi\sim10^{-6}\,\mathrm{m}$.
We emphasize that this choice is not intended as a precise prediction;
it merely serves to examine whether the present framework admits a
self-consistent parameter regime.

Substituting these values into Eq.~\eqref{eq:p:eta}, we obtain
\begin{equation}
\eta
\approx
1.1\times10^{-16}\,m_e
\approx
1.0\times10^{-46}\,\mathrm{kg}.
\end{equation}
Although $\eta$ is extremely small compared with the electron mass,
it sets the scale separating the tail contribution and the light-cone
contribution derived in the present model.
Once $\eta$ is determined, the corresponding value of $\gamma$ follows
directly from Eq.~\eqref{eq:p:ee}.
Since $\gamma$ does not enter the consistency conditions discussed below
explicitly, we do not display its numerical value here.

To verify the consistency of our calculations, we now examine the
dimensionless quantities $\eta t_c$ and $\eta T$.
Restoring $\hbar$ and $c$, the quantity $\eta t_c$ becomes
$\eta t_c c^2/\hbar$ in SI units.
Using the above estimate of $\eta$, we obtain $\frac{\eta t_c c^2}{\hbar}
\approx 2.7\times10^{22} \gg 1$,
confirming that the condition $\eta t_c\gg1$ is well satisfied.

Next, we estimate $\eta T$, which becomes
$\eta T c^2/\hbar$ in SI units.
The parameter $T$ denotes the evolution time over which the
noise-induced phase accumulates.
If this stochastic phase contributes to the gradual loss of quantum
coherence, then $T$ may be interpreted as a characteristic decoherence
timescale.
From Secs.~\ref{sec:pnt} and~\ref{sec:effn}, particularly
Eq.~\eqref{eq:n:Vtex}, the variance of the accumulated phase is
$\frac{y^2\gamma^2 t_c T}{4\pi} = \frac{cT}{\xi}$.
As a rough estimate, we take the accumulated phase to become
statistically significant once its variance reaches a value of order
unity. This corresponds to
$cT/\xi=\mathcal{O}(1)$ and, for the illustrative value
$\xi\sim10^{-6}\,\mathrm{m}$, gives
$T\sim10^{-14}\,\mathrm{s}$.
For such a timescale, $\frac{\eta T c^2}{\hbar} \ll 1$.

In practice, fermions are also subject to interactions with their
environment, such as electromagnetic scattering, which are neglected in
the present analysis.
Our description is therefore applicable only over evolution times
shorter than those on which environmental effects become dominant.
Even adopting the conservative upper estimate
$T\lesssim10^{-5}\,\mathrm{s}$, the condition $\eta T<1$ remains well satisfied.

These illustrative estimates demonstrate that the present framework
admits a broad parameter regime in which the analytical approximations
employed throughout this work are mutually consistent.
A more precise determination of the model parameters would ultimately
require additional theoretical developments and, if the framework proves
physically relevant, confrontation with experimental or observational
constraints.

\section{Summary and outlook}
\label{sec:conclusion}

In this work, we have investigated a stochastic field-theoretic model in which
a complex scalar doublet is coupled to a statistically invariant complex
white-noise field. The resulting random-valued action preserves Lorentz and
local $\mathrm{U}(1)\times\mathrm{SU}(2)$ symmetries at the statistical level,
although individual realizations of the action are not invariant under these
transformations. Consequently, the associated stochastic Euler-Lagrange
equations also possess statistical, rather than deterministic,
$\mathrm{U}(1)\times\mathrm{SU}(2)$ symmetry.

To obtain analytical results, we restricted our analysis to the gauge-field-free
sector by setting the electroweak gauge fields to zero. Within this
approximation, we constructed explicit solutions of the stochastic
Euler-Lagrange equation and found that the scalar-doublet field naturally
separates into two statistically independent contributions. The tail
contribution performs a noise-driven random walk in the internal field space,
leading to a finite expectation value for the magnitude of the scalar doublet,
whereas the light-cone contribution remains statistically independent of the
tail contribution and gives rise to distinct dynamical effects in the Yukawa
sector. The present work should therefore be regarded as a mathematically tractable toy
model for investigating stochastic scalar-doublet dynamics within the
electroweak framework.

We further examined how the stochastic scalar-doublet field influences fermions
through Yukawa interactions. Within the approximations adopted
here, the tail contribution produces an effective fermion mass term, whereas
the light-cone contribution generates a spatially correlated stochastic phase
acting on the fermion wave function. The latter naturally gives rise to a
colored-noise process with a finite correlation length. These results provide a
field-theoretic realization of the phenomenological fermion-colored-noise
coupling introduced in our earlier work~\cite{Wang24}, although the present analysis does not
assume any particular interpretation of the physical role of this noise.

Using the effective fermion mass and the phase-correlation length as
illustrative inputs, we also demonstrated that the model admits a broad
parameter regime in which the analytical approximations employed throughout the
paper are mutually consistent. The numerical estimates presented here should be
viewed only as consistency checks rather than as precise determinations of the
model parameters.

Several important limitations of the present work should be emphasized. First,
the analysis has been restricted to the gauge-field-free sector. Extending the
present formalism to the fully coupled stochastic dynamics of the scalar doublet
and electroweak gauge fields remains an important open problem. Second, the
present model is not intended to reproduce the complete phenomenology of the
Standard Model Higgs sector. In particular, questions concerning the observed
Higgs boson mass, self-interactions, production mechanisms, decay channels, and
scattering processes have not been addressed. Whether these phenomena can be
incorporated into the present stochastic framework remains to be investigated.

The cosmological implications of the model also deserve further study. The
present work demonstrates that the stochastic dynamics remain mathematically
well defined in an expanding FLRW universe and that the scalar-doublet
magnitude evolves dynamically under stochastic driving. A quantitatively
reliable analysis beyond the flat-spacetime approximation will require solving
the tail function $V(x,y)$ in the full FLRW geometry, most likely by numerical
methods. Such calculations will be essential for assessing possible
cosmological consequences of the model and confronting them with observational
constraints.

More broadly, the present work illustrates that replacing deterministic
symmetry by statistical symmetry leads to a mathematically consistent class of
stochastic field theories with qualitatively novel dynamics. Whether such
frameworks can provide useful descriptions of physics beyond the conventional
deterministic formulation of quantum field theory remains an interesting
question for future investigation.

%\bibliography{references}

\end{document}